# Superconductivity of LaH$_{10}$ and LaH$_{16}$ polyhydrides

Ivan A. Kruglov,[1,2,*] Dmitrii V. Semenok,[1,3,*] Hao Song,[9] Radosław Szczęśniak,[4,5] Izabela A. Wrona,[4] Ryosuke Akashi,[6] M. Mahdi Davari Esfahani,[7] Defang Duan,[9] Tian Cui,[9] Alexander G. Kvashnin,[3,1] and Artem R. Oganov [3,2,8,*]

[1] Moscow Institute of Physics and Technology, 9 Institutsky Lane, Dolgoprudny 141700, Russia
[2] Dukhov Research Institute of Automatics (VNIIA), Moscow 127055, Russia
[3] Skolkovo Institute of Science and Technology, Skolkovo Innovation Center, 3 Nobel Street, Moscow 121205, Russia
[4] Institute of Physics, Jan Dlugosz University in Częstochowa, Ave. Armii Krajowej 13/15, 42-200 Częstochowa, Poland
[5] Institute of Physics, Częstochowa University of Technology, Ave. Armii Krajowej 19, 42-200 Częstochowa, Poland
[6] University of Tokyo, 7-3-1 Hongo, Bunkyo, Tokyo 113-8654, Japan
[7] Department of Geosciences and Center for Materials by Design, Institute for Advanced Computational Science, State University of New York, Stony Brook, NY 11794-2100, USA
[8] International Center for Materials Discovery, Northwestern Polytechnical University, Xi'an 710072, China
[9] Key State Laboratory of Superhard Materials, Jilin University, Changchun, China

**Corresponding authors**
*Ivan A. Kruglov, e-mail: ivan.kruglov@phystech.edu
*Dmitrii V. Semenok, e-mail: dmitrii.semenok@skoltech.ru
*Artem. R. Oganov, e-mail: a.oganov@skoltech.ru

**ABSTRACT.** Recent experiments have established previously predicted LaH$_{10}$ as the highest-temperature superconductor, with $T_C$ up to 250–260 K [1]. In this work we explore the high-pressure phase stability and superconductivity of lanthanum hydrides LaH$_m$. We predict the stability of the hitherto unreported polyhydride $P6/mmm$-LaH$_{16}$ at pressures above 150 GPa; at 200 GPa, its predicted superconducting $T_C$ is 156 K, the critical field $\mu_0 H_C(0)$ is approximately 35 T, and the superconducting gap is up to 35 meV. We revisit the superconductivity of LaH$_{10}$ and find its $T_C$ to be up to 259 K at 170 GPa from solving the Eliashberg equation and 271 K from solving the gap equation in the density functional theory for superconductors (SCDFT), which also allows us to compute the Coulomb pseudopotential $\mu^*$ for LaH$_{10}$ and LaH$_{16}$. The presence of several polymorphic modifications of LaH$_{10}$ may explain the variety in its measured $T_C$ values [1, 2].

## Introduction

Recently, superhydrides of various elements attracted a great attention: they were first predicted and then experimentally proven to exhibit a record high-temperature superconductivity under a high pressure. According to the Bardeen–Cooper–Schrieffer (BCS) theory, the metallic hydrogen is expected to be a high-$T_C$ superconductor [3–6], yet the pressure needed for its formation is too high (~ 500 GPa) [7]. Ashcroft proposed to stabilize states similar to metallic hydrogen at lower pressures by creating hydrogen-rich hydrides, where the hydrogen atoms experience an additional "chemical" pressure [8]. Following this, many theoretical predictions of remarkable high-temperature superconductors (HTSC) were made, e.g., in the Ca-H [9], Y-H [10], H-S [11], Th-H [12], Ac-H [13], and La-H [10] systems. In all these systems, the high-$T_C$ superconductors have unusual chemical compositions, like H$_3$S, LaH$_{10}$, or CaH$_6$. The formation of such compounds, violating the rules of classical chemistry, commonly takes place under pressure [15].

The first experimental proof was obtained for H$_3$S with measured $T_C$ = 203 K at 155 GPa [16] (the predicted value was 191–204 K at 200 GPa [11]). The confidence in such predictions has greatly increased because of their agreement with the observations, stimulating a wave of theoretical studies of superconductivity in hydrides [10, 12, 13, 16–29]. The recent theoretical work on the La-H and Y-H systems [10] under pressure showed that at 300 GPa LaH$_{10}$ and YH$_{10}$ may display room-temperature superconductivity (at 286–326 K). According to that theoretical study, at pressures below 200 GPa LaH$_{10}$ has the $R\bar{3}m$ symmetry, which changes to $Fm\bar{3}m$ at higher pressures. In the experiments by Geballe et al. [20], $R\bar{3}m$-LaH$_{10}$ was found at ≤ 160 to 170 GPa, while $Fm\bar{3}m$-LaH$_{10}$ was seen at higher pressures. The experimentally observed sharp drop in resistivity in the LaH$_{10}$ samples was the first evidence of superconductivity in this system [1, 2]. Different studies showed a series of superconducting-like transitions with $T_C$ of 70 and 112 K [1], 210–215 K [1], 244–250 K [1, 2], and 260–280 K [1]. This discrepancy in the $T_C$ values may be caused by the coexistence of lanthanum hydrides having different compositions and crystal structures in the studied samples. To address these questions, we predicted the stable compounds in the La-H system using the variable-composition evolutionary algorithm USPEX [30–32]. While most studies report only $T_C$ and assume $\mu^* = 0.1$–0.15 for the calculations, here we obtain $T_C$ using various approaches including the superconducting DFT, which requires no assumption of $\mu^*$, and also compute other properties, such as the critical magnetic field [1].

## Computational Details

The evolutionary algorithm USPEX [30–32] is a powerful tool for predicting thermodynamically stable compounds of given elements. To predict thermodynamically stable phases in the La-H system, we performed variable-composition crystal structure searches at pressures from 50 to 200 GPa. The first generation of 100 structures was created using a random symmetric [32] and topological structure generators [33] with the number of atoms in the primitive unit cell ranging from 8 to 16, while the subsequent generations contained 20% of random structures, and the rest 80% of structures were created using the heredity, softmutation, and transmutation operators. Here, the evolutionary searches were combined with structure relaxations using the density functional theory (DFT) [34, 35] within the generalized gradient approximation (the Perdew–Burke–Ernzerhof functional) [36], and the projector-augmented wave method [37, 38] as implemented in the VASP package [39–41]. The plane wave kinetic energy cutoff was set to 500 eV and the Brillouin zone was sampled by Γ-centered $k$-points meshes with a resolution of $2\pi \times 0.05$ Å$^{-1}$. This methodology was used and proved to be very effective in our previous works (e.g., [12–14]).

The calculations of the critical temperature and electron-phonon coupling (EPC) parameters were carried out using Quantum ESPRESSO (QE) package [42] within the density functional perturbation theory [43], employing the plane-wave pseudopotential method and Perdew–Burke–Ernzerhof exchange-correlation functional [36]. The convergence tests showed that 90 Ry is a suitable kinetic energy cutoff for the plane wave basis set. The electronic band structures and phonon densities of states were calculated using both VASP (the finite displacement method using PHONOPY [44, 45]) and QE (the density functional perturbation theory [43]), which demonstrated a good consistency.

In our calculations of the EPC parameter λ, the first Brillouin zone was sampled using a 4×4×4 $q$-points mesh and a denser 16×16×16 $k$-points mesh (with the Gaussian smearing and σ = 0.02 Ry, which approximates the zero-width limits).

For LaH$_{10}$ and LaH$_{16}$, we also estimated $T_C$ by solving the gap equation within the density functional theory for superconductors (SCDFT) [46, 47]:

$$\Delta_{nk}(T) = -Z_{nk}(T)\Delta_{nk}(T) - \frac{1}{2}\sum K_{nkn'k'}(T)\frac{tanh\beta E_{n'k'}}{E_{n'k'}}\Delta_{n'k'}(T). \quad (1)$$

Here, $E_{nk} = \sqrt{[\Delta_{nk}(T)]^2 + [\xi_{nk}]^2}$, with $\xi_{nk}$ being the normal-state energy eigenvalue of $\hat{H}_e$ labeled by the band index $n$ and wave vector $k$ measured from the Fermi level. The "order parameter" $\Delta_{nk}(T)$ depends on $n$ and $k$, not on the frequency ω; it is defined in a way different from that in the Eliashberg equation and is proportional to the thermal average $\langle c_{nk\uparrow}c_{n-k\downarrow}\rangle$, with $c_{nk\sigma}$ being the annihilation operator of the spin state $nk\sigma$ [46]. The interaction effects treated in the Eliashberg equation are included with the kernels entering this gap equation; $Z_{nk} = Z_{nk}^{ph}$; $K_{nkn'k'} = K_{nkn'k'}^{ph} + K_{nkn'k'}^{el}$. Here we included the mass-renormalization by the phonon exchange with $Z_{nk}^{ph} = Z^{ph}(\xi_{nk})$ (see Eq. (40) in [48] and an improved form of Eq. (24) in [47] to consider the nonconstant electronic density of states), the phonon-mediated electron-electron attraction with $K^{PH}_{nkn'k'} = K^{PH}(\xi_{nk}, \xi_{n'k'})$ (Eq. (23) in [47]), and the screened Coulomb repulsion $K^{el}_{nkn'k'}$ (Eq. (3) in [53]). Unlike the Eliashberg equation, the SCDFT gap equation does not contain the frequency ω. Nevertheless, the retardation effect [50] is approximately incorporated [47]. The absence of ω enables us to treat all electronic states in a wide energy range of ±30 eV at an affordable computational cost, and therefore estimate $T_C$ without the empirical Coulomb pseudopotential µ*. We estimated $T_C$ as the temperature where the nontrivial solution of this gap equation vanishes (see Supporting Information [51]).

The Coulomb pseudopotential µ*, included in the Eliashberg equation and its approximate solutions such as the Allen–Dynes (A-D) equation, quantifies the impact of the renormalized Coulomb repulsion. Its value can in principle be estimated with the matrix element of the screened Coulomb interaction and energy scales of the electrons and phonons [50, 51], but practically it is determined by fitting the calculated observables with the experimental data or simply by accepting the typical values (e.g., ~ 0.10–0.13) from such previous fitting results [53]. On the other hand, while the SCDFT gives us the framework free from empirical parameters such as µ*, deriving from it some interesting observables is difficult compared with the Eliashberg theory. Especially, the spectral gap at absolute zero, being defined within the Green's function framework, is not directly accessible within the SCDFT. To estimate several observable quantities, we took a hybrid approach reconciling these frameworks: Determining the value of µ* so that the Eliashberg (or Allen–Dynes) equation reproduces $T_C$ derived from the SCDFT gap equation [54] and then employing it in later calculations within the Eliashberg theory. In this approach, we presume that the retardation effect [50] can be expressed by a single parameter, µ*; possible drawbacks of this method are discussed later.

In this work we did not consider the effect of anharmonicity on the dynamic and superconducting properties of lanthanum hydrides. While this effect can be important, previous studies showed that the superconducting properties predicted within the "conventional" approach are in relatively good agreement with the experimental data.

## Results and Discussion

### I. Stability of La-H phases

We performed the variable-composition evolutionary searches for stable compounds and crystal structures at 50, 100, 150, and 200 GPa. The thermodynamic convex hulls are shown in Fig. 1. By composition, stable phases are those that appear on the convex hull. Thus, at 50 GPa we found $Fm\bar{3}m$-LaH, $Pnma$-LaH$_3$, $Cmc2_1$-LaH$_7$, and $Cc$-LaH$_9$ to be stable (Fig. 1a). At 100 GPa, LaH$_7$ and LaH$_9$ disappear from the convex hull, while four new hydrides become stable: $Cm$-LaH$_2$, $Cmcm$-LaH$_3$, $P\bar{1}$-LaH$_5$, and $P4/nmm$-LaH$_{11}$ (Fig. 1b). At 150 GPa, the chemistry of lanthanum hydrides is much richer: $P6/mmm$-LaH$_2$, $Cmcm$-LaH$_3$, $I4/mmm$-LaH$_4$, $P\bar{1}$-LaH$_5$, $Fm\bar{3}m$-LaH$_{10}$, and $P6/mmm$-LaH$_{16}$ phases become stable (Fig. 1c). At 200 GPa, only five stable lanthanum hydrides remain on the convex hull: $P6/mmm$-LaH$_2$, $Cmmm$-La$_3$H$_{10}$, $I4/mmm$-LaH$_4$, $Fm\bar{3}m$-LaH$_{10}$, and $P6/mmm$-LaH$_{16}$. Our results are similar to those from [55], the main difference being that we predict the new LaH, La$_3$H$_{10}$, and LaH$_{16}$ compounds and also find that LaH$_{10}$ is stable at pressures above 135 GPa, which is consistent with the experimental data [1, 2]. Yet, we found $Fm\bar{3}m$-LaH$_{10}$ to be more stable than $R\bar{3}m$-LaH$_{10}$ at pressures above 150 GPa. The $R\bar{3}m \rightarrow Fm\bar{3}m$ phase transition occurs at 128 GPa (Supporting Information Fig. S19 [51]). Taking into account the zero-point energy correction, the pressure of this phase transition shifts to 135 GPa, which is consistent with the experimental phase transition pressure of 160 GPa [20]. Below we discuss the superconducting properties of the previously known lanthanum hydrides and explore the crystal structure and SC properties of the newly found LaH$_{16}$. The crystal structure data are summarized in Supporting Information Table S1 [51].

Stoichiometry 1:1 is common for various hydrides (e.g., U-H [14], Fe-H [23], and many others [56]). $Fm\bar{3}m$-LaH has a structure of a rock salt type, with La–H distance of 2.26 Å at 50 GPa. In the *Pnma*-LaH$_3$ structure, the La atoms have a 10-fold coordination and H–H distances are too long to be considered bonding (2.28 Å at 50 GPa). In *Cmc2$_1$*-LaH$_7$, the lanthanum atoms have a 13-fold coordination, while the shortest H–H distances (0.81 Å at 50 GPa) are clearly bonding and quite close to the H–H bond length in the molecular hydrogen (0.74 Å).

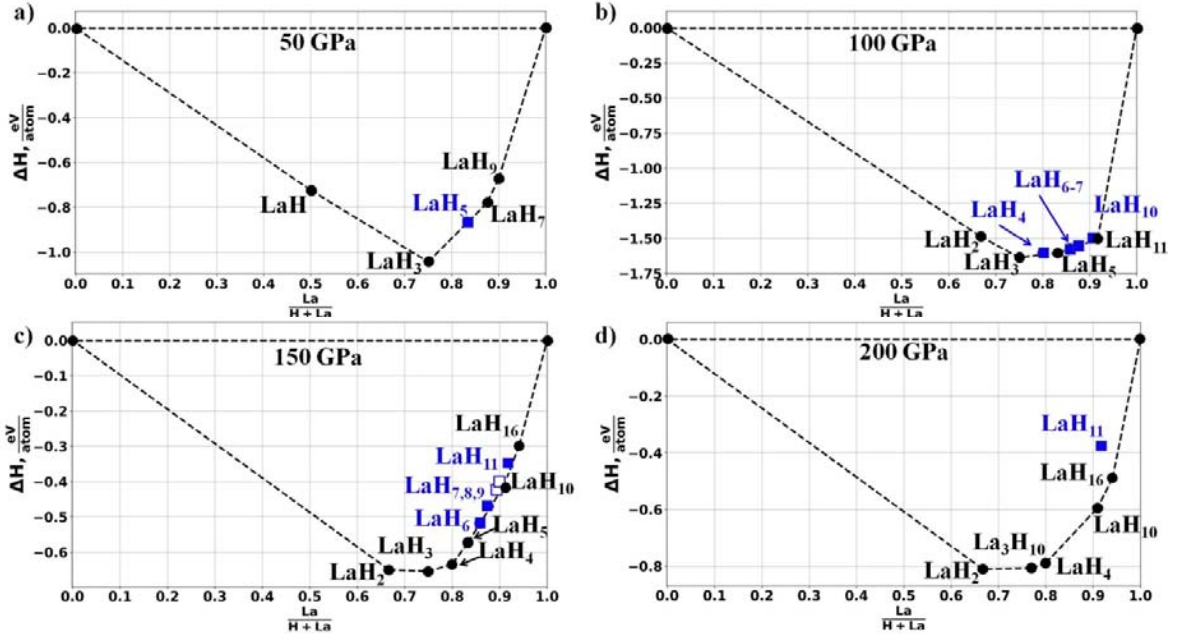

Fig. 1. Convex hull diagrams for the La-H system at a) 50, b) 100, c) 150, and d) 200 GPa. The metastable and stable phases are marked by blue squares and black circles, respectively.

Our results at 150 GPa are consistent with those in [10]: *P6/mmm*-LaH$_2$, *Cmcm*-LaH$_3$, *Immm*-LaH$_4$, and *P$\bar{1}$*-LaH$_5$ phases are thermodynamically stable. Previous studies explored only simple fixed stoichiometries, whereas the USPEX method used here can automatically detect even nontrivial stoichiometries, which helped us find a new phase, *Cmmm*-La$_3$H$_{10}$, at 200 GPa. This phase is structurally related to *Cmcm*-LaH$_3$, has an additional hydrogen atom in the tripled unit cell, and its lattice parameter *c* is 3 times larger.

Superconducting properties for the most promising LaH$_{10}$ and LaH$_{16}$ phases were calculated using the SCDFT method, then the $\mu^*$ value was determined by adjusting $T_C$ obtained through the numerical solution of the Eliashberg equation to that from the SCDFT. For comparison, the $\mu^*$ values determined using the A-D formula are presented in Supporting Information Table S2 [51]. We find a significant difference in the $\mu^*$ values obtained, depending on whether the Eliashberg or A-D equation were used, because the LaH$_x$ systems are classified as the strong-coupling case where these equations give substantially different values of $T_C$ [1].

## II. Superconductivity of LaH$_{10}$ phases

Superconductivity of LaH$_{10}$ has been confirmed in experiments [1, 2]. Our analysis is based on the $T_C$ calculation using the SCDFT, Eliashberg function $\alpha^2 F(\omega)$, and the electronic density of states for *R$\bar{3}$m*- and *Fm$\bar{3}$m*-LaH$_{10}$ phases. The experimental observations found the maximum $T_C \sim 251$ K at 168 GPa [2] or $T_C \sim 260$ K at 180 GPa [1], which is close to the *R$\bar{3}$m*–*Fm$\bar{3}$m* phase transition (but in the cubic phase). The value of the Coulomb pseudopotential $\mu^*$ for the LaH$_{10}$ phases determined using the SCDFT equals 0.2 at 200 GPa (see Supporting Information [51]).

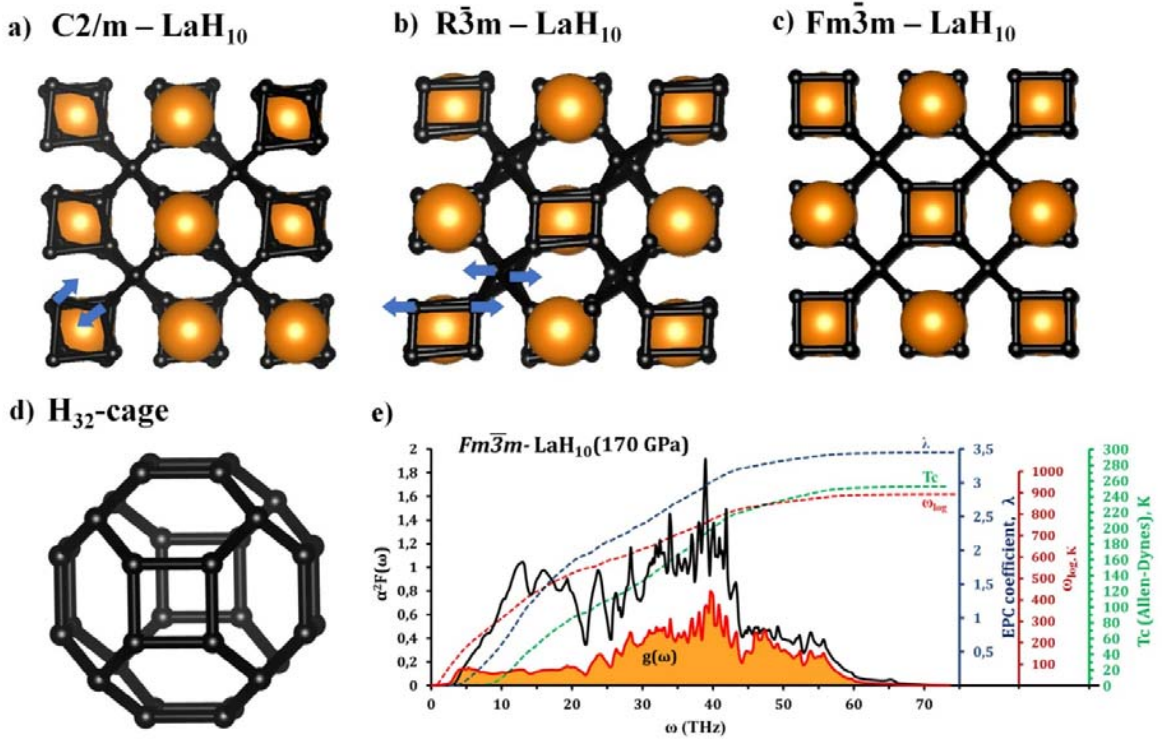

Fig. 2. Crystal structures of a) $C2/m$-LaH$_{10}$, b) $R\bar{3}m$-LaH$_{10}$, c) $Fm\bar{3}m$-LaH$_{10}$, and d) hydrogen H$_{32}$ cage common for these phases; e) the Eliashberg $\alpha^2F(\omega)$ function (black curve), phonon density of states (orange), cumulative $\omega_{log}$ (red), EPC coefficient $\lambda$ (blue), and critical transition temperature $T_C$ (green) of $Fm\bar{3}m$-LaH$_{10}$ at 170 GPa. The crystal structures were generated using VESTA software [57]; the atoms of La and hydrogen are shown as large orange and small black balls, respectively; blue arrows show displacements of the hydrogen atoms in comparison with the cubic H$_8$ cluster.

For $Fm\bar{3}m$-LaH$_{10}$, the calculations within the DFPT give a very high EPC coefficient $\lambda = 3.75$, which leads to $T_C = 271$ K at 200 GPa (see Table 1 and Supporting Information for details [51]). At 200 and 250 GPa, the calculated volumes (30.4 and 28.5 Å$^3$, respectively) and the electronic density of states, $N(E_F) = N(0) = 10.6$ and 10.0 states/f.u./Ry, allowed us to estimate the Sommerfeld constant (Table 1) at 0.016 and 0.011 J/mol·K$^2$, which is very close to that of $Fm\bar{3}m$-ThH$_{10}$ (0.011 J/mol·K$^2$ [12]) at 100 GPa. These constants were used to calculate the critical magnetic field ($\mu_0H_C(0)$) and the jump in specific heat $\Delta C/T_C$ (Table 1). The calculation details are presented in Supporting Information [51].

The Eliashberg function for the cubic $Fm\bar{3}m$-LaH$_{10}$ at 200 GPa is shown in Fig. 2e (for other pressure values and phases, see Supporting Information [51]). The numerical solution of the Eliashberg equation for the cubic LaH$_{10}$ at the experimental pressure of 170 GPa results in $T_C = 259$ K (Table 1), which is in close agreement with the theoretical results [10] and experiments [1, 2].

The calculated critical temperature for $R\bar{3}m$-LaH$_{10}$ at 150 GPa is 203 K (Table 1), much lower than that of the $Fm\bar{3}m$ modification, which can explain the experimental results of Drozdov et al. [1].

**Table 1.** Parameters of the superconducting state of $Fm\bar{3}m$-LaH$_{10}$ and $R\bar{3}m$-LaH$_{10}$ from the Eliashberg equation (with $\mu^* = 0.2$).

| Parameter | $Fm\bar{3}m$-LaH$_{10}$ | | | | $R\bar{3}m$-LaH$_{10}$ | |
|---|---|---|---|---|---|---|
| | 170 GPa | 200 GPa | 210 GPa | 250 GPa | 150 GPa | 165 GPa |
| $N_f$, states/f.u./Ry | 11.2 | 10.6 | 10.3 | 10.0 | 12.2 | 11.0 |
| $\lambda$ | 3.94 | 3.75 | 3.42 | 2.29 | 2.77 | 2.63 |
| $\omega_{\log}$, K | 801 | 906 | 851 | 1253 | 833 | 840 |
| $T_C$, K | 259 | 271* | 249 | 246 | 203 | 197 |
| $\Delta(0)$, meV | 62.0 | 63.7 | 59.6 | 48.0 | 48.5 | 43.7 |
| $\mu_0 H_C(0)$, T** | 89.0 | 95.0 | 81.0 | 66.7 | 72.7 | 71.0 |
| $\Delta C/T_C$, mJ/mol·K$^2$ | 31.5 | 44.7 | 34.5 | 33.4 | 42.8 | 25.7 |
| $\gamma$, J/mol·K$^2$ | 0.019 | 0.016 | 0.015 | 0.011 | 0.018 | 0.018 |
| $R_\Delta = 2\Delta(0)/k_B T_C$ | 5.54 | 5.46 | 5.55 | 5.00 | 5.55 | 5.54 |

\* SCDFT (see Supporting Information [51])
\*\* The experimentally measured critical magnetic field is 95–136 T [1]

The isotope coefficient β was calculated using the A-D formula (see Supporting Information [51]). With $\mu^* = 0.11$ used for the A-D equation, the calculated value was the same, β = 0.48, for both $Fm\bar{3}m$-LaH$_{10}$ and $R\bar{3}m$-LaH$_{10}$ at 200 GPa. Using the isotope coefficient β, the critical temperature of lanthanum decadeuteride can be determined as $T_D = 2^{-\beta} \cdot T_H$, resulting in 181 K for $Fm\bar{3}m$-LaD$_{10}$ and 154 K for $R\bar{3}m$-LaD$_{10}$. The experimentally measured $T_C$ for $Fm\bar{3}m$-LaD$_{10}$ is 168 K [1], corresponding with our prediction.

### III. Prediction of high-$T_C$ LaH$_{16}$

Perhaps the most intriguing finding of our work is the prediction of the thermodynamic stability of $P6/mmm$-LaH$_{16}$ at pressures above 150 GPa (Fig. 1c). This phase has a crystal structure different from the previously predicted AcH$_{16}$ [13]. The lanthanum atoms form an *hcp* sublattice and are coordinated by 12 hydrogen atoms (Fig. 3a). These H-units consist of three parallel layers of hydrogen atoms. The distances between the hydrogen atoms in the first, second, and third layers at 150 GPa are 1.07, 1.02, and 1.07 Å, respectively. The distance between the closest hydrogen atoms from different layers is 1.25 Å, so these layers form infinite 2D networks. The crystal structures of all predicted phases are summarized in Supporting Information Table S1 [51].

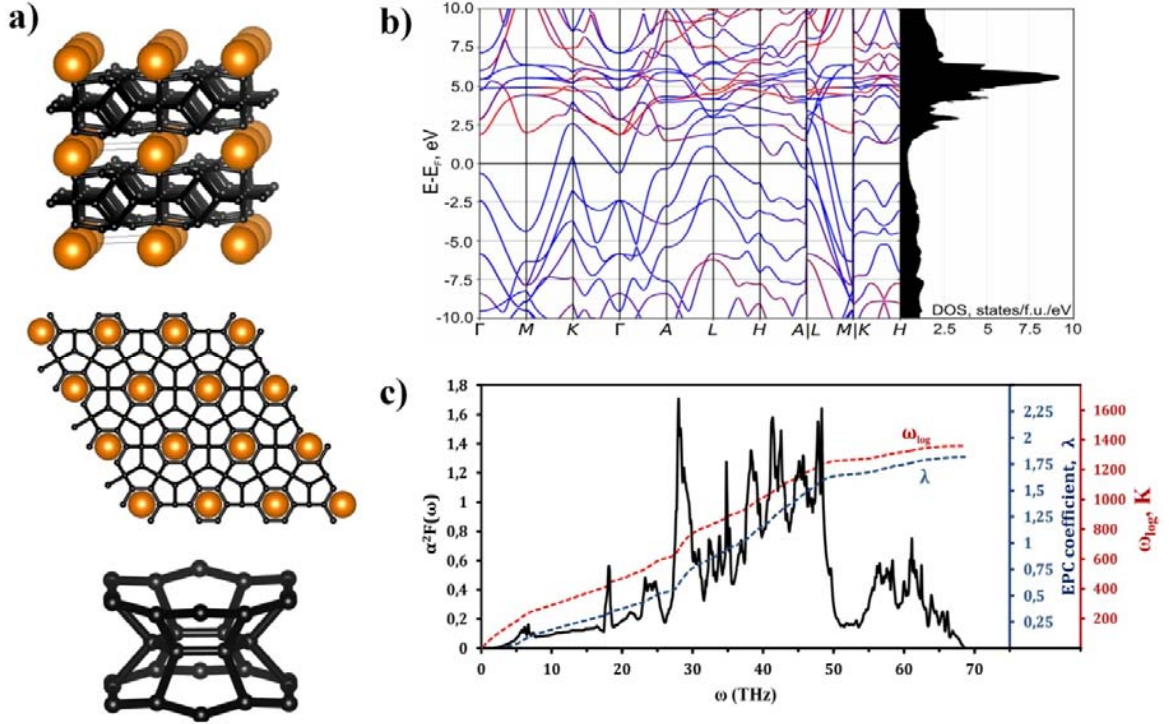

Fig. 3. a) Crystal structure of $P6/mmm$-LaH$_{16}$ at 150 GPa; b) band structure and DOS of LaH$_{16}$ (red lines are for La, blue — for H) at 250 GPa; c) the Eliashberg function $\alpha^2F(\omega)$ (black), cumulative $\omega_{\log}$ (red), and the EPC coefficient $\lambda$ (blue) of $P6/mmm$-LaH$_{16}$ at 200 GPa.

$P6/mmm$-LaH$_{16}$ shows very little dependence of the electronic density of states on the pressure at 200-300 GPa. The Eliashberg function of $P6/mmm$-LaH$_{16}$ computed at 200 GPa is shown in Fig. 3c (for other pressure values, see Supporting Information [51]). The EPC coefficient of LaH$_{16}$ (Table 2) is approximately two times lower than that of LaH$_{10}$. The SCDFT calculations for LaH$_{16}$ at 200 GPa yield $T_C = 156$ K, which leads to the calculated $\mu^* = 0.41$, an anomalously large value compared to the commonly accepted 0.1–0.15 interval. This abnormal value of $\mu^*$ for LaH$_{16}$ is caused by the complex behavior of DOS($E$) in the vicinity of $E_F \pm 1$ eV (see Supporting Information [51]). We used $\mu^* = 0.41$ in the numerical solution of the Eliashberg equation to calculate the superconducting properties of $P6/mmm$-LaH$_{16}$ (Table 2). $T_C$ of this phase shows a weak pressure dependence ($dT_C/dP = -0.3$ K/GPa).

**Table 2.** Parameters of the superconducting state of $P6/mmm$-LaH$_{16}$ at 200–300 GPa calculated using the Eliashberg equation with $\mu^* = 0.41$. Here $\gamma$ is the Sommerfeld constant calculated using Eq. *Ошибка! Источник ссылки не найден.*.

| Parameter | 200 GPa | 250 GPa | 300 GPa |
|---|---|---|---|
| $N_f$, states/f.u./Ry | 7.21 | 6.94 | 7.07 |
| $\lambda$ | 1.82 | 1.63 | 1.44 |
| $\omega_{\log}$, K | 1362 | 1511 | 1675 |
| $T_C$, K | 156 | 141 | 118 |
| $\Delta(0)$, meV | 29.6 | 25.7 | 20.2 |
| $\mu_0 H_C(0)$, T | 35.0 | 29.9 | 23.6 |
| $\Delta C/T_C$, mJ/mol·K$^2$ | 18.6 | 15.1 | 12.4 |
| $\gamma$, mJ/mol·K$^2$ | 7.3 | 6.5 | 6.1 |
| $R_\Delta = 2\Delta(0)/k_B T_C$ | 4.5 | 4.2 | 4.0 |

Table 3 shows the critical temperatures for LaH$_{10}$ and LaH$_{16}$ calculated using the SCDFT and the values of $\mu^*$ adjusted so that the Eliashberg and A-D equations give the same values of $T_C$.

**Table 3.** Superconducting parameters of LaH$_{10}$ and LaH$_{16}$ phases

| Parameter | LaH$_{10}$ at 200 GPa | LaH$_{16}$ at 200 GPa |
| --- | --- | --- |
| Critical temperature ($T_C$), K | 271 | 156 |
| $\mu^*$ (Eliashberg eq.) | 0.20 | 0.41 |
| $\mu^*$ (Allen–Dynes eq.) | 0.11 | 0.198 |

## IV. Superconductivity of LaH$_m$ ($m$ = 4, …, 9, 11)

It was shown before that different lanthanum hydrides can undergo a series of superconducting transitions at different pressures with different $T_C$ [1], so we checked other lanthanum hydrides to explore this possibility. In this work, we found a number of LaH$_m$ phases ($m$ = 4, …, 9, 11) at different pressures, and calculating their $T_C$ using the SCDFT method or the numerical solution of the Eliashberg equation would have been very computationally expensive. Therefore, we calculated $T_C$ of LaH$_m$ phases ($m$ = 4, …, 9, 11) using the Allen–Dynes (A-D) formula [52, 57] in the pressure range of 150–180 GPa, which is the experimentally studied pressure region [1], with the commonly accepted $\mu^*$ values (0.1–0.15). We consider this interval of $\mu^*$ reasonable because for LaH$_{10}$ and LaH$_{16}$ the A-D formula yielded $\mu^*$ equal to 0.11 and 0.2, respectively (Supporting Information Table S2 [51]). The calculated $T_C$ values are shown in Fig. 4. The results point to LaH$_4$, LaH$_6$, and LaH$_7$ as the compounds that could undergo the experimentally observed transition at ~ 215 K with the critical magnetic field $\mu_0 H_C(0)$ = 60–70 T; the other transition at ~ 112 K with the critical magnetic field $\mu_0 H_C(0)$ = 20–25 T could occur in LaH$_5$, LaH$_8$, and LaH$_{11}$. More details about the calculation of $T_C$ in these phases can be found in Supporting Information [51].

## V. Possible departure from the standard Eliashberg picture

In the present analysis, we estimated the superconducting parameters with the A-D formula and isotropic (constant-DOS) Eliashberg equation. They are based on a simple standard view of the detailed electronic structure such as the constant electronic DOS and screened Coulomb matrix elements. The degree of departure of the target system from this simple view can be estimated by analyzing the Coulomb pseudopotential $\mu^*$ [54]. We evaluated the nonrenormalized $\mu$ as the Fermi surface average of the screened Coulomb interaction [50, 51], which was 0.18 and 0.11 for LaH$_{10}$ and LaH$_{16}$, respectively. The straightforward use of the renormalization formula
$$\mu^* = \mu/[1 + \mu\ln(E_{el}/\omega_D)], \qquad (2)$$
where $E_{el}$ is the Fermi level calculated from the band bottom and $\omega_D$ is the Debye frequency, yields $\mu^*$ = 0.10 and 0.07, which is far smaller than our estimate from the SCDFT results (0.20 and 0.41, respectively). The deviation of $\mu^*$ estimated using the SCDFT approach from the empirically accepted range 0.10–0.13 [50, 52] has been pointed out in the previous SCDFT studies [53, 58]. It is thought to indicate the substantial effects ignored in Eq. (2), such as the energy dependence of the electronic DOS and the Kohn–Sham state dependence of the Coulomb matrix element, which are incorporated in the SCDFT framework. Notably, we found a paradoxical relation, $\mu < \mu^*$, derived from the SCDFT, which has also been observed in H$_3$S [60]. The DOS effect would be accounted for more properly by the DOS-dependent Eliashberg equation [61]. However, the single-parameter approximation of the Coulomb effect is likely less reliable in hydrides because of the large phonon energy scale comparable to the electronic one. The ab initio treatment of the Coulomb matrix element in the Eliashberg equation [62] would reveal such effects on the superconducting parameters evaluated above, which is out of the scope of the present study.

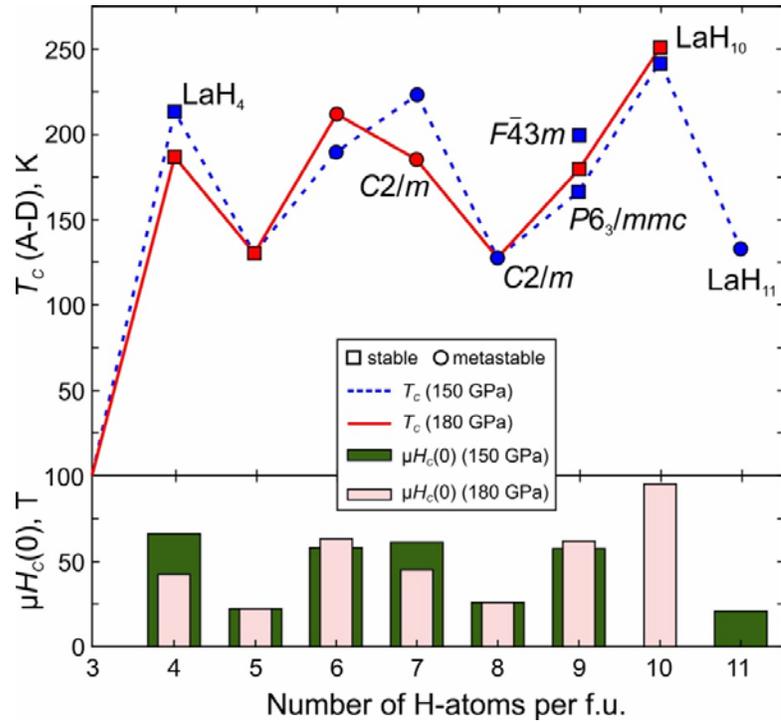

Fig. 4. Superconducting temperatures $T_C$, obtained using the A-D formula with $\mu^* = 0.1$, and critical magnetic fields for a series of lanthanum hydrides at 150 and 180 GPa.

## Conclusions

Using the ab initio evolutionary algorithm USPEX, we predicted stable superconducting compounds in the La-H system at 50, 100, 150, and 200 GPa, including the previously unknown polyhydride $P6/mmm$-LaH$_{16}$. The electronic, phonon, and superconducting properties of LaH$_m$ with $m = 4, …, 11$ were studied within the harmonic approximation, which gives possible explanations of the observed superconducting transitions at 70, 112, and 215 K [1]. The SCDFT calculations for LaH$_{10}$ and LaH$_{16}$ led to unusually high values of $\mu^*$ (0.2–0.41) for the numerical solution of the Eliashberg equation, lying far outside the conventionally accepted interval, 0.1–0.15. The Eliashberg function calculated for $Fm\bar{3}m$-LaH$_{10}$ at 170 GPa yields $T_C = 259$ K, according to the numerical solution of the Eliashberg equation with $\mu^* = 0.2$. The critical magnetic field of $Fm\bar{3}m$-LaH$_{10}$ was found to be 89–95 T (73 T for the $R\bar{3}m$ phase), the harmonic isotope coefficient $\beta = 0.48$. The values of all the superconducting parameters for lanthanum hydrides are in close agreement with the available experimental data [1]. We found that $P6/mmm$-LaH$_{16}$ is a high-temperature superconductor with $T_C$ of up to 156 K, the critical magnetic field of ~ 35 T, and the superconducting gap of 30 meV at 200 GPa. The main conclusion of this investigation is that even in one system the $\mu^*$ value, especially for the numerical solution of the Eliashberg equation, may significantly vary and go beyond the commonly accepted interval, 0.1–0.15, which lends a special importance to the SCDFT calculations.

## Acknowledgments

A.R.O. acknowledges the support by the Russian Science Foundation (grant 19-72-30043). A.G.K. thanks the RFBR project № 19-03-00100 and FASIE for the financial support within the UMNIK grant № 13408GU/2018. The calculations were performed on Rurik supercomputer at MIPT and Arkuda supercomputer of the Skolkovo Foundation.

# Supporting Information

# Superconductivity in LaH$_{10}$ and LaH$_{16}$ polyhydrides

*Ivan A. Kruglov,* [1,2,*] *Dmitrii V. Semenok,* [1,3] *Hao Song* [9], *Radosław Szczęśniak,* [4,5] *Izabela A. Wrona,* [4] *Ryosuke Akashi,* [6] *M. Mahdi Davari Esfahani,* [7] *Defang Duan,* [9] *Tian Cui,* [9] *Alexander G. Kvashnin* [3,1] *and Artem R. Oganov* [3,2,8]

[1] Moscow Institute of Physics and Technology, 141700, 9 Institutsky lane, Dolgoprudny, Russia
[2] Dukhov Research Institute of Automatics (VNIIA), Moscow 127055, Russia
[3] Skolkovo Institute of Science and Technology, Skolkovo Innovation Center 143026, 3 Nobel Street, Moscow, Russia
[4] Institute of Physics, Jan Dlugosz University in Czestochowa, Ave. Armii Krajowej 13/15, 42-200 Częstochowa, Poland
[5] Institute of Physics, Częstochowa University of Technology, Ave. Armii Krajowej 19, 42-200 Częstochowa, Poland
[6] University of Tokyo, 7-3-1 Hongo, Bunkyo, Tokyo 113-8654, Japan
[7] Department of Geosciences and Center for Materials by Design, Institute for Advanced Computational Science, State University of New York, Stony Brook, NY 11794-2100, USA
[8] International Center for Materials Discovery, Northwestern Polytechnical University, Xi'an, 710072, China
[9] Key State Laboratory of Superhard Materials, Jilin University, Changchun, China

**Corresponding Author**
*Ivan A. Kruglov, E-mail: ivan.kruglov@phystech.edu
*Dmitrii V. Semenok, E-mail: Dmitrii.Semenok@skoltech.ru
*Artem. R. Oganov, E-mail: A.Oganov@skoltech.ru

**Content**

# Content

# Crystal data of La-H phases

**Table S1**. Crystal structures of predicted La-H phases.

| Phase | Pressure, GPa | Lattice parameters | | Coordinates | | |
|---|---|---|---|---|---|---|
| $Fm\bar{3}m$-LaH | 50 | a = 4.53 Å | La | 0.5 | 0.5 | 0.5 |
| | | | H | 0.0 | 0.0 | 0.0 |
| $Pnma$-LaH$_3$ | 50 | a = 7.20 Å, b = 4.57 Å, c = 3.67 Å | La | -0.365 | 0.25 | -0.320 |
| | | | H | 0.394 | -0.001 | -0.171 |
| | | | H | 0.356 | 0.25 | 0.332 |
| $Cmc2_1$-LaH$_7$ | 50 | a = 7.20 Å, b = 4.57 Å, c = 3.67 Å | La | 0.0 | -0.159 | -0.452 |
| | | | H | 0.0 | 0.466 | 0.408 |
| | | | H | 0.0 | 0.245 | -0.300 |
| | | | H | 0.0 | -0.365 | 0.241 |
| | | | H | 0.0 | 0.174 | 0.420 |
| | | | H | 0.25 | 0.417 | 0.237 |
| | | | H | 0.0 | -0.497 | -0.465 |
| $P6/mmm$-LaH$_2$ | 150 | a = 2.80 Å, c = 2.72 Å | La | 0.0 | 0.0 | 0.0 |
| | | | H | 0.333 | 0.667 | 0.5 |
| $Cmcm$-LaH$_3$ | 150 | a = 2.76 Å, b = 10.69 Å, c = 2.80 Å | La | 0.0 | 0.118 | 0.25 |
| | | | H | 0.0 | 0.467 | 0.25 |
| | | | H | 0.0 | 0.313 | 0.25 |
| | | | H | 0.0 | -0.25 | 0.25 |
| $Cmmm$-La$_3$H$_{10}$ | 150 | a = 16.70 Å, b = 2.77 Å, c = 2.75 Å | La | -0.326 | 0.0 | 0.5 |
| | | | La | 0.0 | 0.0 | 0.0 |
| | | | H | -0.089 | 0.0 | 0.5 |
| | | | H | 0.128 | 0.0 | 0.0 |
| | | | H | -0.228 | 0.0 | 0.0 |
| | | | H | -0.449 | 0.0 | 0.5 |
| | | | H | 0.411 | 0.0 | 0.0 |
| $I4/mmm$-LaH$_4$ | 150 | a = 2.74 Å, c = 6.03 Å | La | 0.0 | 0.0 | 0.5 |
| | | | H | 0.0 | 0.5 | 0.25 |
| | | | H | 0.0 | 0.0 | -0.153 |
| $P\bar{1}$-LaH$_5$ | 150 | a = 2.91 Å, b = 5.26 Å, c = 3.47 Å, α = 93.34, β = 110.24, γ = 98.71 | La | -0.327 | -0.266 | -0.273 |
| | | | H | 0.263 | 0.036 | -0.255 |
| | | | H | 0.073 | 0.431 | -0.280 |
| | | | H | -0.067 | 0.122 | -0.349 |
| | | | H | -0.304 | 0.375 | -0.108 |
| | | | H | -0.376 | 0.112 | -0.151 |
| $R\bar{3}m$-LaH$_{10}$ | 150 | a = 3.66 Å, c = 8.53 Å | La | 0.0 | 0.0 | 0.5 |
| | | | H | 0.0 | 0.0 | 0.097 |
| | | | H | -0.168 | 0.168 | 0.276 |
| | | | H | 0.0 | 0.0 | -0.259 |
| $Fm\bar{3}m$-LaH$_{10}$ | 150 | a = 5.08 Å | La | 0.0 | 0.0 | 0.0 |
| | | | H | -0.378 | -0.378 | -0.378 |
| | | | H | 0.25 | 0.25 | 0.25 |
| $P6/mmm$-LaH$_{16}$ | 150 | a = 3.68 Å, c = 3.70 Å | La | 0.0 | 0.0 | 0.0 |
| | | | H | -0.276 | 0.0 | 0.5 |
| | | | H | 0.5 | 0.0 | 0.245 |
| | | | H | 0.333 | 0.667 | -0.203 |

# Equations for calculating $T_C$ and related parameters

The superconducting transition temperature was calculated by solving the gap equation derived from the density functional theory for superconductors (SCDFT) [1,2], where no empirical parameter such as µ* was used. The *nk*-averaged formula for $K^{ph}$ [Eq.(23) in Ref. [2]] and $Z^{ph}$ [Eq.(40) in Ref. [3]] were employed to use the calculated $α^2F(ω)$. The nk-dependent formula for $K^{ph}$ [Eq.(3) in Ref. [4]] was used, which includes the dynamical as well as static screening effect on the Coulomb repulsion. The dielectric matrices $ε(iω)$ were represented with the auxiliary plane-wave energy cutoff of 12.8 Ry and were calculated within the random-phase approximation [5] and electron-electron Coulomb matrix elements were evaluated as

$$W_{nk,n'k'}(i\omega) = <\psi_{nk}\psi_{n-k} | \varepsilon^{-1}(i\omega)V | \psi_{n'k'}\psi_{n'-k'}> \quad (S1)$$

where *V* is being the bare Coulomb interaction. Low-energy Kohn-Sham states were accurately treated with the weighted random sampling scheme in solving the gap equation, as described in Ref. [2]. The Kohn-Sham energy eigenvalues and matrix elements for the sampled states were evaluated by the linear interpolation from the data on equal meshes. The detailed conditions are summarized in Table S2.

**Table S2.** Summary of the detailed conditions for the Tc calculation with the SCDFT scheme.

| Class | Parameter | LaH$_{10}$ 200GPa | LaH$_{16}$ 200GPa |
|---|---|---|---|
| charge density | k | 16x16x16 equal mesh | 16x16x16 equal mesh |
| | interpolation | 1st order Hermite Gaussian [5] with width 0.050Ry | |
| dielectric matrix | k for bands crossing $E_F$ | 15x15x15 equal mesh | 15x15x15 equal mesh |
| | k for other bands | 5x5x5 equal mesh | 5x5x5 equal mesh |
| | number of unoccupied bands [a] | 57 | 75 |
| | interpolation | Tetrahedron with the Rath-Freeman treatment [6] | |
| DOS for phononic kernels | k | 21x21x21 equal mesh | 21x21x21 equal mesh |
| | interpolation | Tetrahedron with the Blöchl correction [7] | |
| SCDFT gap function | number of unoccupied bands [b] | 24 | 36 |
| | k for the electronic kernel | 5x5x5 equal mesh | 5x5x5 equal mesh |
| | k for the KS energies | 21x21x21 equal mesh | 21x21x21 equal mesh |
| | Sampling points for bands crossing $E_F$ | 6000 | 6000 |
| | Sampling points for the other bands | 200 | 200 |
| | Sampling error in Tc (%) | ~2.6 | ~1.9 |
| SC parameters | Critical temperature ($T_C$) | 271 | 156 |
| | Coloumb pseudopotential µ* (E) µ* (Allen-Dynes) | 0.20 0.11 | 0.41 0.198 |

[a] States up to $E_F$+70eV are taken into account.
[b] States up to $E_F$+40eV are taken into account.

Isotopic coefficient $β_{AD}$ was calculated using the Allen-Dynes interpolation formulas:

$$\beta_{McM} = -\frac{d \ln T_C}{d \ln M} = \frac{1}{2}\left[1 - \frac{1.04(1+\lambda)(1+0.62\lambda)}{[\lambda - \mu^*(1+0.62\lambda)]^2}\mu^{*2}\right] \tag{S2}$$

$$\beta_{AD} = \beta_{McM} - \frac{2.34\mu^{*2}\lambda^{3/2}}{(2.46+9.25\mu^*)\cdot((2.46+9.25\mu^*)^{3/2}+\lambda^{3/2})} - \frac{130.4\cdot\mu^{*2}\lambda^2(1+6.3\mu^*)\left(1-\frac{\omega_{\log}}{\omega_2}\right)\frac{\omega_{\log}}{\omega_2}}{\left(8.28+104\mu^*+329\mu^{*2}+2.5\cdot\lambda^2\frac{\omega_{\log}}{\omega_2}\right)\cdot\left(8.28+104\mu^*+329\mu^{*2}+2.5\cdot\lambda^2\left(\frac{\omega_{\log}}{\omega_2}\right)^2\right)} \tag{S3}$$

where the last two correction terms are usually small (~0.01).

The critical temperature of superconducting transition was calculated using Matsubara-type linearized Eliashberg equations: [8]

$$\hbar\omega_j = \pi(2j+1)\cdot k_B T, \quad j = 0, \pm 1, \pm 2, \ldots \tag{S4}$$

$$\lambda(\omega_i - \omega_j) = 2\int_0^\infty \frac{\omega \cdot \alpha^2 F(\omega)}{\omega^2 + (\omega_i - \omega_j)^2} d\omega \tag{S5}$$

$$\Delta(\omega = \omega_i, T) = \Delta_i(T) = \pi k_B T \sum_j \frac{[\lambda(\omega_i - \omega_j) - \mu^*]}{\rho + \left|\hbar\omega_j + \pi k_B T \sum_k (\text{sign}\,\omega_k)\cdot\lambda(\omega_j - \omega_k)\right|}\cdot\Delta_j(T) \tag{S6}$$

where $T$ is temperature in Kelvins, $\mu^*$ is Coloumb pseudopotential, $\omega$ is frequency in Hz, $\rho(T)$ is a pair-breaking parameter, function $\lambda(\omega_i - \omega_j)$ relates to effective electron-electron interaction *via* exchange of phonons. [9] Transition temperature can be found as a solution of equation $\rho(T_C) = 0$ where $\rho(T)$ is defined as $max(\rho)$ providing that $\Delta(\omega)$ is not a zero function of $\omega$ at fixed temperature.

These equations can be rewritten in a matrix form as [10]

$$\rho(T)\psi_m = \sum_{n=0}^N K_{mn}\psi_n \Leftrightarrow \rho(T)\begin{pmatrix}\psi_1\\ \ldots \\ \psi_N\end{pmatrix} = \begin{pmatrix}K_{11} & \ldots & K_{1N}\\ \ldots & K_{ii} & \ldots \\ K_{N1} & \ldots & K_{NN}\end{pmatrix}\times\begin{pmatrix}\psi_1\\ \ldots \\ \psi_N\end{pmatrix}, \tag{S7}$$

where $\psi_n$ relates to $\Delta(\omega, T)$, and

$$K_{mn} = F(m-n) + F(m+n+1) - 2\mu^* - \delta_{mn}\left[2m + 1 + F(0) + 2\sum_{l=1}^m F(l)\right] \tag{S8}$$

$$F(x) = F(x,T) = 2\int_0^{\omega\max}\frac{\alpha^2 F(\omega)}{(\hbar\omega)^2 + (2\pi\cdot kT\cdot x)^2}\cdot\hbar^2\omega d\omega, \tag{S9}$$

where $\delta_{nn} = 1$ and $\delta_{nm} = 0$ ($n \neq m$) – is a unit matrix. Now one can replace criterion of $\rho(T_C) = 0$ by the vanishing of the maximum eigenvalue of the matrix $K_{nm}$: $\{\rho = \text{max\_eigenvalue}(K_{nm}) = f(T), f(T_C) = 0\}$.

We have implemented this simple approach, proposed by Philip Allen in 1974, as Matlab functions, shown at the last page of Supporting Information, which read files with $\alpha^2F$ functions and carry out the calculations of $T_C$.

Exact calculations for *P6/mmm*-LaH$_{16}$ at 250 GPa were also made using the Eliashberg equations in the following form (on the imaginary axis) [8]:

$$\phi_n = \frac{\pi}{\beta} \sum_{m=-M}^{M} \frac{\lambda(i\omega_n - i\omega_m) - \mu^* \theta(\omega_c - |\omega_m|)}{\sqrt{(\hbar \omega_m Z_m)^2 + \phi_m^2}} \phi_m \tag{S10}$$

$$Z_n = 1 + \frac{1}{\omega_n} \frac{\pi}{\beta} \sum_{m=-M}^{M} \frac{\lambda(i\omega_n - i\omega_m)}{\sqrt{(\hbar \omega_m Z_m)^2 + \phi_m^2}} \omega_m Z_m \tag{S11}$$

where: $\phi_n$ is the order parameter function, $Z_n$ is the wave function renormalization factor, $\theta(x)$ is Heaviside function, $\hbar\omega_n = \pi \cdot k_B T \cdot (2n-1)$ is the *n*-th Matsubara frequency, $\beta = 1/k_B T$, $\mu^*$ is Coulomb pseudopotential, $\omega_c$ is the cut-off energy ($\omega_c = 3\Omega_{max}$, $\Omega_{max}$ is the maximum frequency in $\alpha^2F(\omega)$). The electron-phonon pairing kernel is defined as:

$$\lambda(z) = 2 \int_0^{\Omega_{max}} \frac{\alpha^2 F(\Omega)}{\Omega^2 - z^2} \Omega \, d\Omega. \tag{S12}$$

The Eliashberg equations were solved for 2201 Matsubara frequencies, starting from $T_0 = 10$ K (below this temperature, the solutions are not stable). The methods discussed in the papers [11–13] were used during the calculations.

The full form of the order parameter and the wave function renormalization factor on the real axis was obtained through analytical extension of the Eliashberg equations solutions from the imaginary axis, using the formula:

$$X(\omega) = \frac{p_1 + p_2 \omega + \ldots + p_r \omega^{r-1}}{q_1 + q_2 \omega + \ldots + q_r \omega^{r-1} + \omega^r}, \tag{S13}$$

where $X \in \{\Delta; Z\}$, $r = 50$. The values of $p_j$ and $q_j$ parameters were selected in accordance with the principles presented in work [14]. Obtained results allow one to calculate dimensionless parameter $R_C$ (relative jump of the specific heat), using the formula below:

$$R_C = \frac{C^S(T_C) - C^N(T_C)}{C^N(T_C)} \tag{S14}$$

Because of strong-coupling and retardation effects in LaH$_{10}$ and LaH$_{16}$, parameter $R_C$ differs significantly from BCS theory prediction, in which the constant value is equal to 1.43. [15]

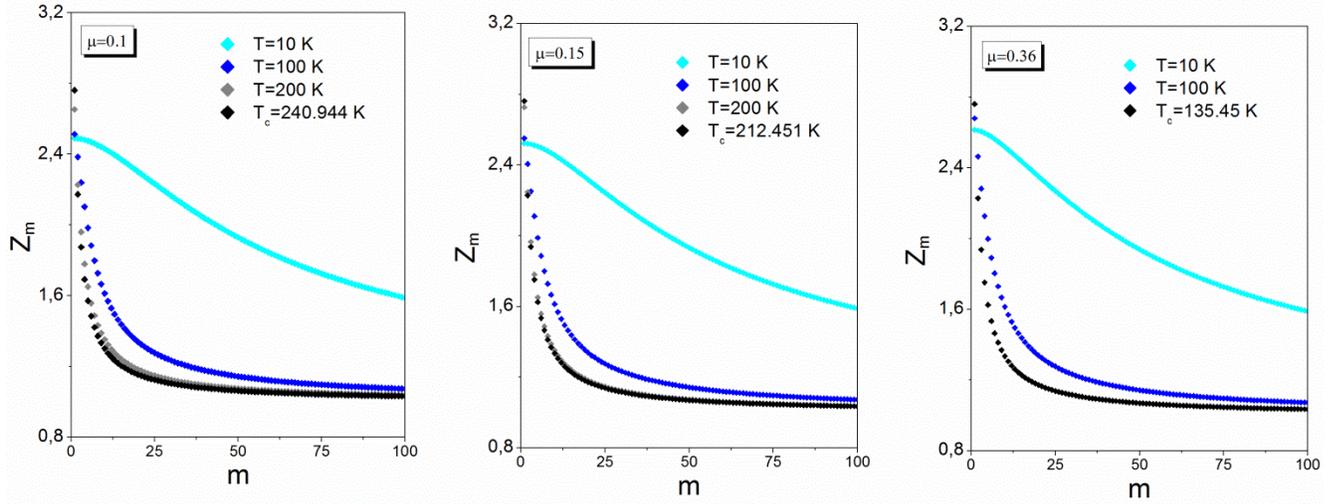

Fig. S1. The wave function renormalization factor for *P6/mmm*-LaH$_{16}$ at 250 GPa on the imaginary axis for the selected values of temperature (the first 100 values).

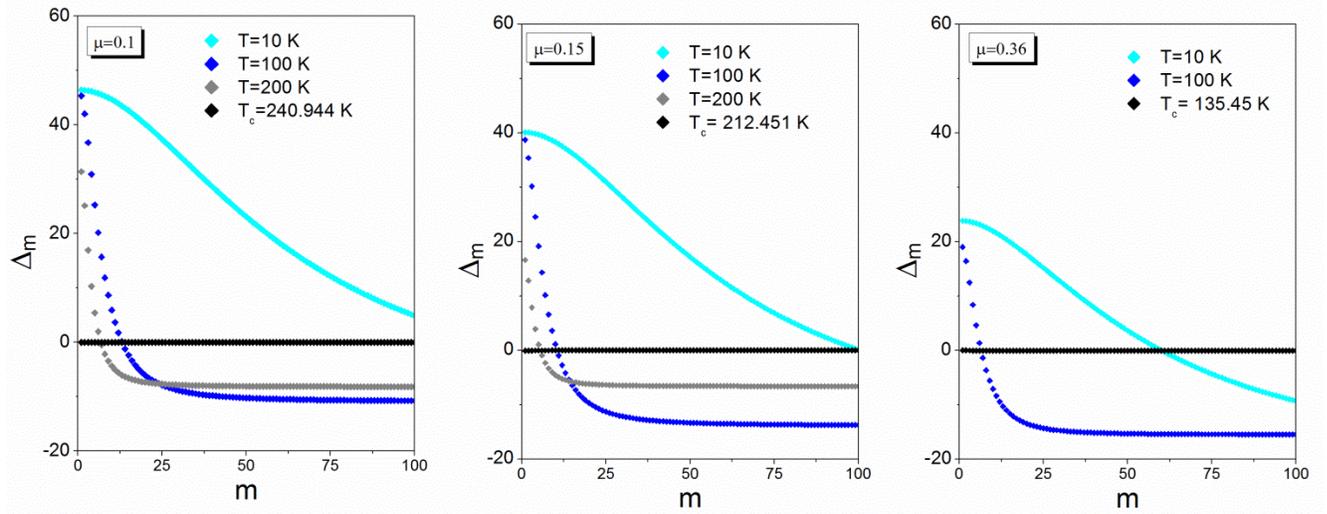

Fig. S2. The order parameter (in meV) for *P6/mmm*-LaH$_{16}$ at 250 GPa on the imaginary axis for the selected values of temperature (the first 100 values) and Coloumb pseudopotential.

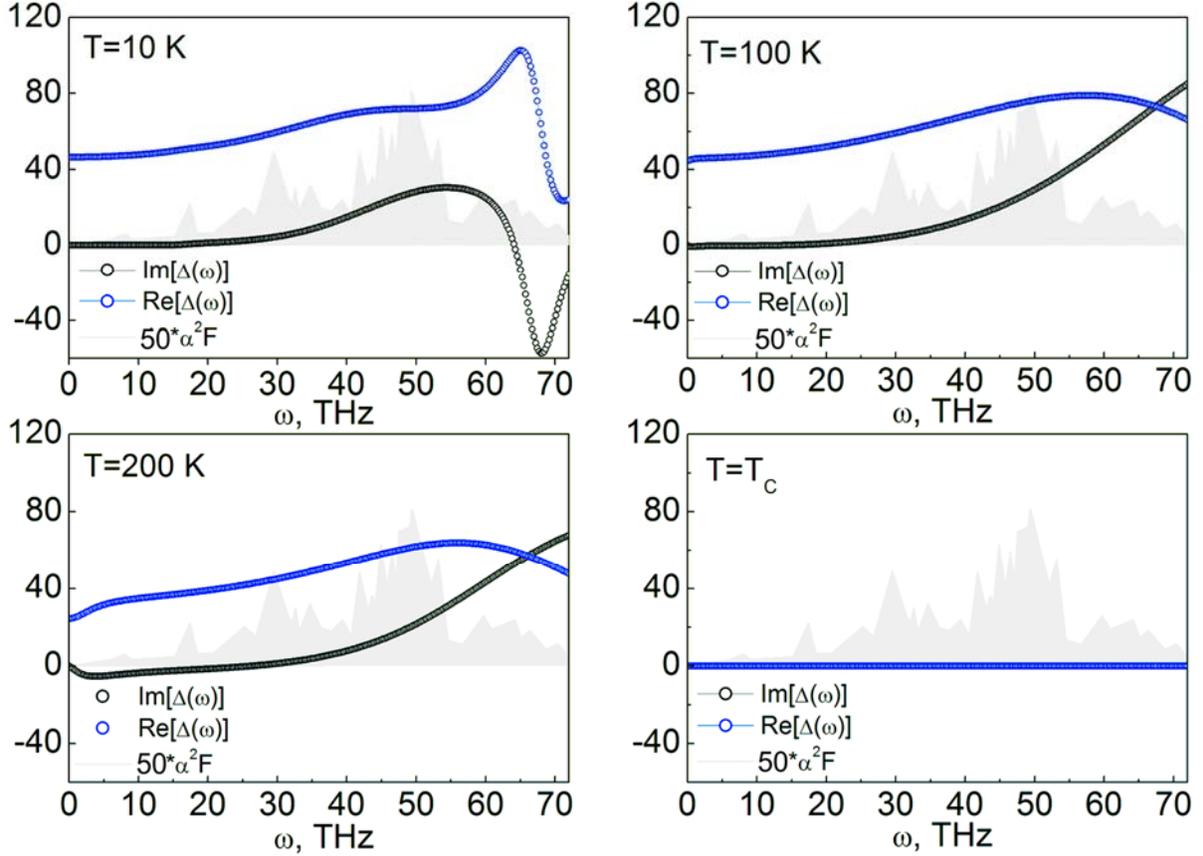

Fig. S3. The order parameter $\Delta(\omega)$ in meV for $P6/mmm$-LaH$_{16}$ at 250 GPa on the real axis for the selected values of temperature ($\mu^* = 0.1$). In addition, the Eliashberg function (grey) was added as the background (multiplied by 50 for easier comparison).

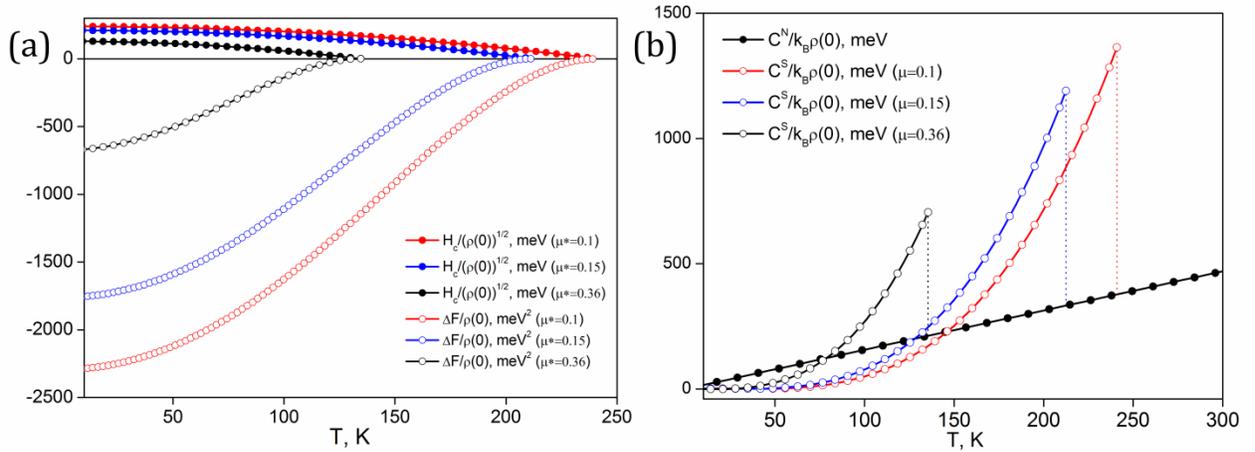

Fig. S4. a) The dependence of the thermodynamic critical field on the temperature (in meV, solid circles) and free energy difference between the superconducting state and the normal state as a function of the temperature (in meV$^2$, open circles); b) The specific heat of the superconducting state ($C_S$) and the normal state ($C_N$) as a function of temperature. Here $\rho(0)$ is the density of states at the Fermi level.

**Table S3**. Parameters of superconductive state of $P6/mmm$-LaH$_{16}$ at 250 GPa calculated by the Eliashberg equations at different $\mu^*$

| Parameter | $P6/mmm$-LaH$_{16}$ (250 GPa) | | |
|---|---|---|---|
| | $\mu^*=0.1$ | $\mu^*=0.15$ | $\mu^*=0.36$ |
| $T_C$, K | 241 | 213 | 135 |
| $\Delta(0)$, meV | 46.4 | 40.1 | 23.9 |
| $\mu_0 H_C(0)$, T | 52.3 | 45.8 | 28.4 |
| $\Delta C/T_C$, mJ/mol·K$^2$ | 17.3 | 17.2 | 15.4 |
| $\gamma$, mJ/mol·K$^2$ | 6.3 | 6.3 | 6.3 |
| $R_\Delta = 2\Delta(0)/k_B T_C$ | 4.47 | 4.38 | 4.09 |

The Sommerfeld constant was calculated using the following formula:

$$\gamma = \frac{2}{3}\pi^2 k_B^2 N(0)(1+\lambda), \quad (S15)$$

where $\lambda$ is the EPC parameter, $N(0)$ – the electronic density of states at the Fermi level, k$_B$ – the Boltzmann constant.

## Calculation of critical magnetic field for LaH$_{10}$ phases

Using semi-empirical formulas of BCS theory (see Ref. [16], eq. 5.11 and 5.9):

$$\frac{\gamma T_C^2}{H_C^2(0)} = 0.168\left[1 - 12.2\left(\frac{T_C}{\omega_{\log}}\right)^2 \ln\left(\frac{\omega_{\log}}{3T_C}\right)\right] \tag{S16}$$

and

$$\frac{\Delta C(T_C)}{\gamma T_C} = 1.43\left[1 + 53\left(\frac{T_C}{\omega_{\log}}\right)^2 \ln\left(\frac{\omega_{\log}}{3T_C}\right)\right], \tag{S17}$$

we estimated the critical magnetic field and the jump in specific heat for $Fm\bar{3}m$-LaH$_{10}$ in harmonic approximation (see Table 2). The superconducting gap was estimated by well-known semi-empirical equation, working satisfactorily for $T_C/\omega_{\log} < 0.25$ (see equation 4.1 in Ref. [16]):

$$\frac{2\Delta(0)}{k_B T_C} = 3.53\left[1 + 12.5\left(\frac{T_C}{\omega_{\log}}\right)^2 \ln\left(\frac{\omega_{\log}}{2T_C}\right)\right] \tag{S18}$$

Estimation of the critical magnetic field for $Fm\bar{3}m$-LaH$_{10}$ gives $\mu_0 H_C(0) \sim$ 89-95 T at 170-200 GPa, which is greater than 70 T for H$_3$S [17] and agrees with lower bound of recent experimental results of 95-136 T [18].

## Superconductivity of lanthanum hydrides LaH$_m$ ($m$ = 4-9, 11)

According to Ref. [18], pure LaH$_3$ is not a superconductor at all investigated pressures (0-200 GPa), but higher hydrides of La are metallic and superconducting. Superconducting properties of all predicted hydrides were computed by Allen-Dynes (A-D) formula [19,16] at the pressure range 150-180 GPa, which is the experimentally studied pressure region [18].

Surprisingly, $I4/mmm$-LaH$_4$ demonstrates ~~very~~ relatively strong electron-phonon coupling (see Table S4). At 150 GPa the $T_C$(A-D) is found to be 87 K and it decreases to 67 K at 180 GPa with $dT_C/dP \approx -0.67$ K/GPa. Thus, $I4/mmm$-LaH$_4$ could explain superconducting transitions with $T_C$ = 70-112 K observed in Ref. [18].

$P\bar{1}$-LaH$_5$ is stable at pressures above 150 GPa (Fig. 1c) and has $T_C$ of 104-130 K at 150 GPa (see Table S4), almost insensitive to pressure ($dT_C/dP \approx -0.033$ K/GPa). This could explain the superconducting transition observed at $T_C \sim 112$ K [18].

$R\bar{3}m$-LaH$_6$ is metastable, but at 150 GPa it is very close to the convex hull (only 6 meV/atom above it, see Fig. 1c). Its electronic density of states is approximately pressure-independent, and its $T_C$ increases with pressure ($dT_C/dP \approx +0.73$ K/GPa), because of the increase of the characteristic phonon frequency ~~with pressure~~. According to our calculations, $T_C$ of this phase can reach 211 K at 180 GPa (see Table S4).

**Table S4.** Predicted superconducting properties of slightly metastable and stable lanthanum hydrides. $\mu^*$ was taken as 0.1 (0.15).

| Phase | P, GPa | $\lambda$ | $N_f$, states/f.u./Ry | $\omega_{log}$, K | $T_C$ (A-D), K | $\mu_0 H_C(0)$, T |
|---|---|---|---|---|---|---|
| $I4/mmm$-LaH$_4$ [b] | 150 | 1.28 | 8.43 | 796 | 87 (69) | 25 (22) |
|  | 180 | 0.84 | 7.62 | 1255 | 69 (48) | 16 (13) |
| $P\bar{1}$-LaH$_5$ [b] | 150 | 1.21 | 5.57 [a] | 1307 | 130 (104) | 22 (17) |
|  | 180 | 1.15 |  | 1389 | 129 (102) |  |
| $R\bar{3}m$-LaH$_6$ | 150 | 2.89 | 9.04 [a] | 765 | 189 (163) | 58 (51) |
|  | 180 | 2.60 |  | 956 | 211 (183) | 63 (55) |
| $C2/m$-LaH$_7$ | 150 | 2.94 | 7.21 | 894 | 223 (193) | 61 (54) |
|  | 180 | 1.92 | 7.55 | 1102 | 185 (158) | 45 (38) |
| $C2/m$-LaH$_8$ [c] | 150 | 1.56 | 6.32 [a] | 944 | 128 (107) | 26 (21) |
|  | 180 | 1.53 |  | 950 | 127 (106) |  |
| $P6_3/mmc$-LaH$_9$ | 150 | 2.75 | 11.6 [a] | 702 | 166 (144) | 57 (50) |
|  | 180 | 2.98 |  | 708 | 180 (156) | 62 (55) |
| $F\bar{4}3m$-LaH$_9$ | 150 | 2.92 | 7.72 | 802 | 199 (172) | 56 (49) |
| $P4/nmm$-LaH$_{11}$ | 150 | 1.54 | 3.94 | 986 | 133 (111) | 21 (17) |

a) no changes with pressure; b) stable phase; c) from Ref. [20].

$C2/m$-LaH$_7$ is also metastable (17 meV/atom above the convex hull at 150 GPa, Fig. 1c) and displays $T_C$ up to 223 K (with $dT_C/dP \approx -1.26$ K/GPa), see Table 3. This phase may be responsible for observed transition at ~ 215 K [18]. Metastable $C2/m$-LaH$_8$, proposed in Ref. [20], has $T_C$ = 106-127 K (Table S4), which is almost insensitive to pressure.

For metastable LaH$_9$ we considered two polymorphs: isostructural to recently found $P6_3/mmc$-CeH$_9$ [21] and $F\bar{4}3m$-UH$_9$ [22]. We predict $T_C$ = 144-166 K (150 GPa, Table S4) for hexagonal modification with $dT_C/dP$ = +0.46 K/GPa. Obtained value is higher than that for CeH$_9$ [21,23]. For cubic LaH$_9$ $T_C$ is higher: up to 199 K at 150 GPa (Table S4).

Metastable $P4/nmm$-LaH$_{11}$ at 150-180 GPa shows relatively low $T_C$ = 111-133 K (due to low density of states at Fermi level).

**Electronic and phonon properties of La-H phases**

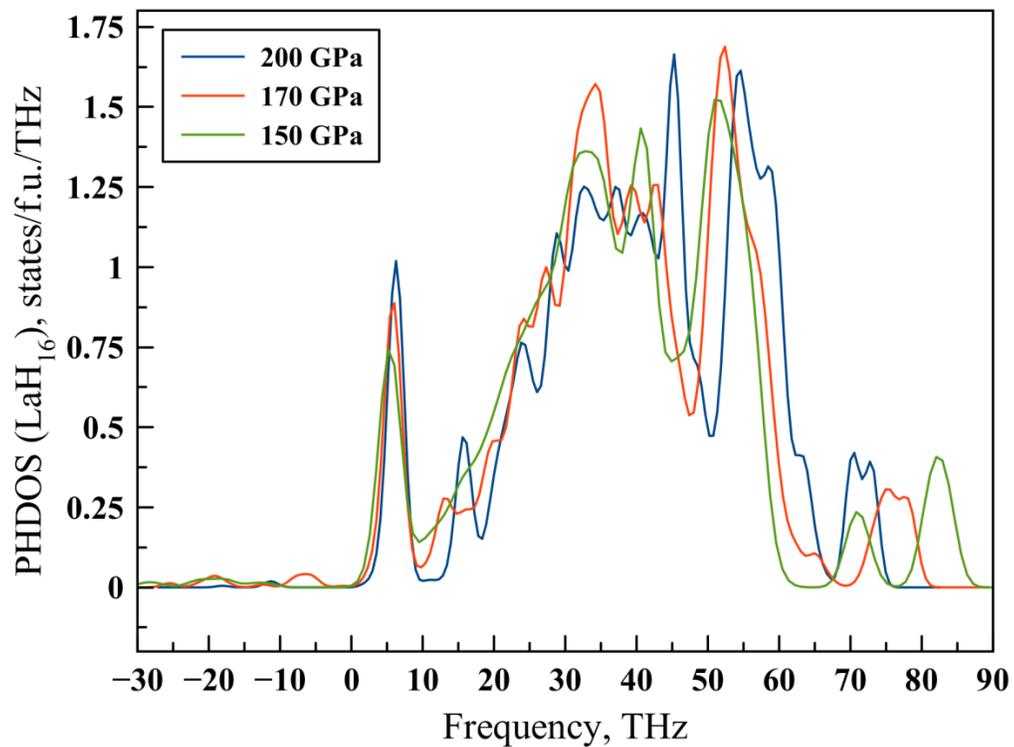

Fig. S5. Phonon density of states (DOS) of *P6/mmm*-LaH$_{16}$ phase at 150, 170 and 200 GPa.

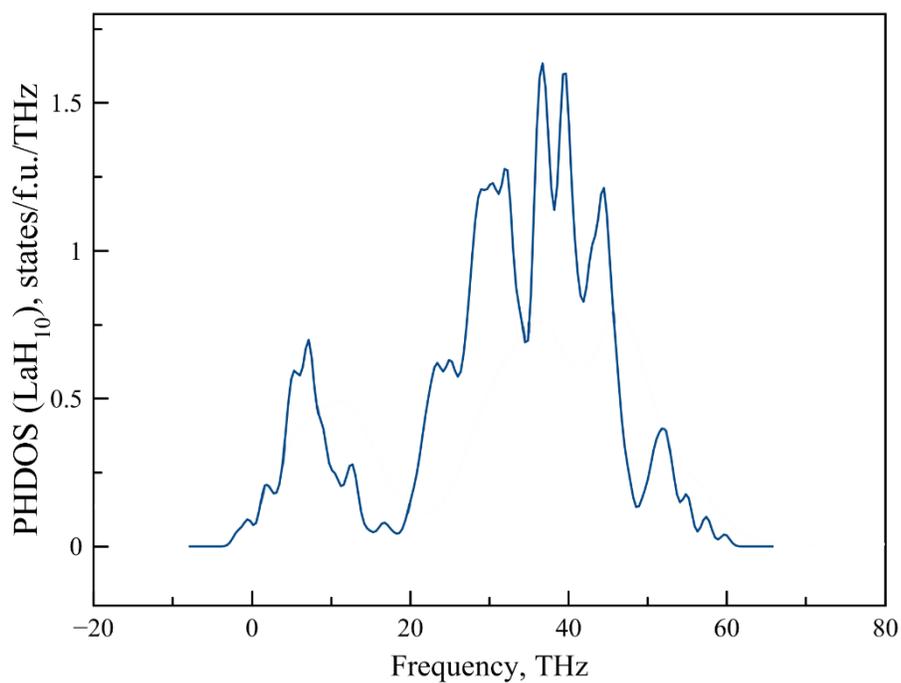

Fig. S6. Phonon density of states (smoothed) of $Fm\bar{3}m$-LaH$_{10}$ at 150 GPa.

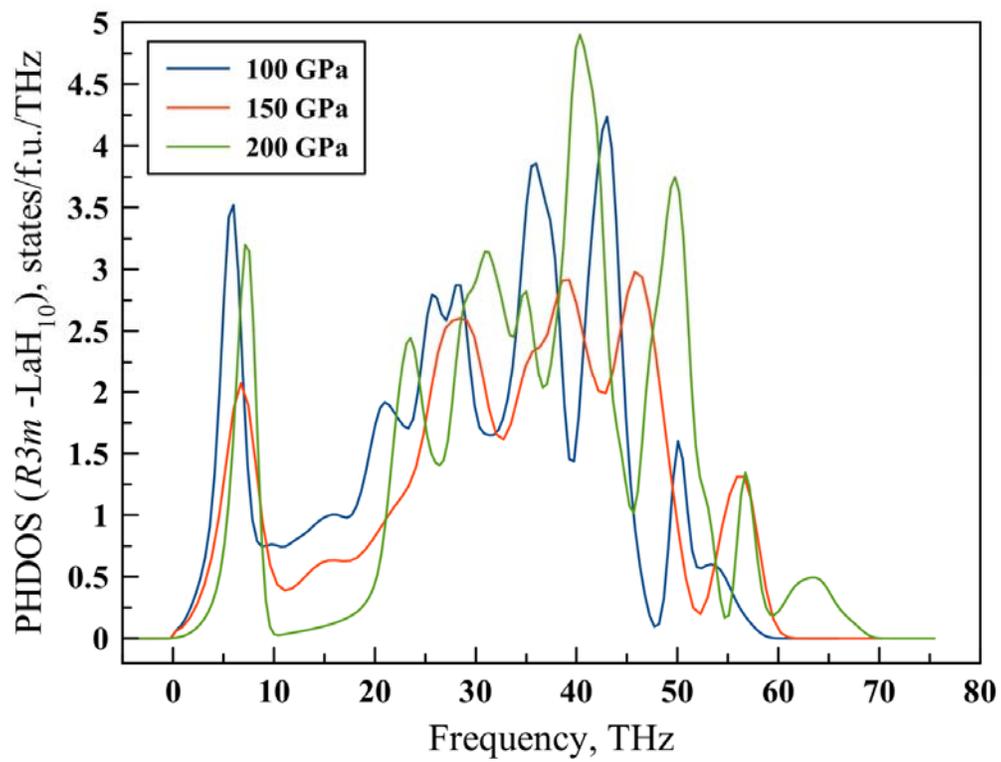

Fig. S7. Phonon density of states (smoothed) of $R\bar{3}m$-$LaH_{10}$ at 100, 150 and 200 GPa.

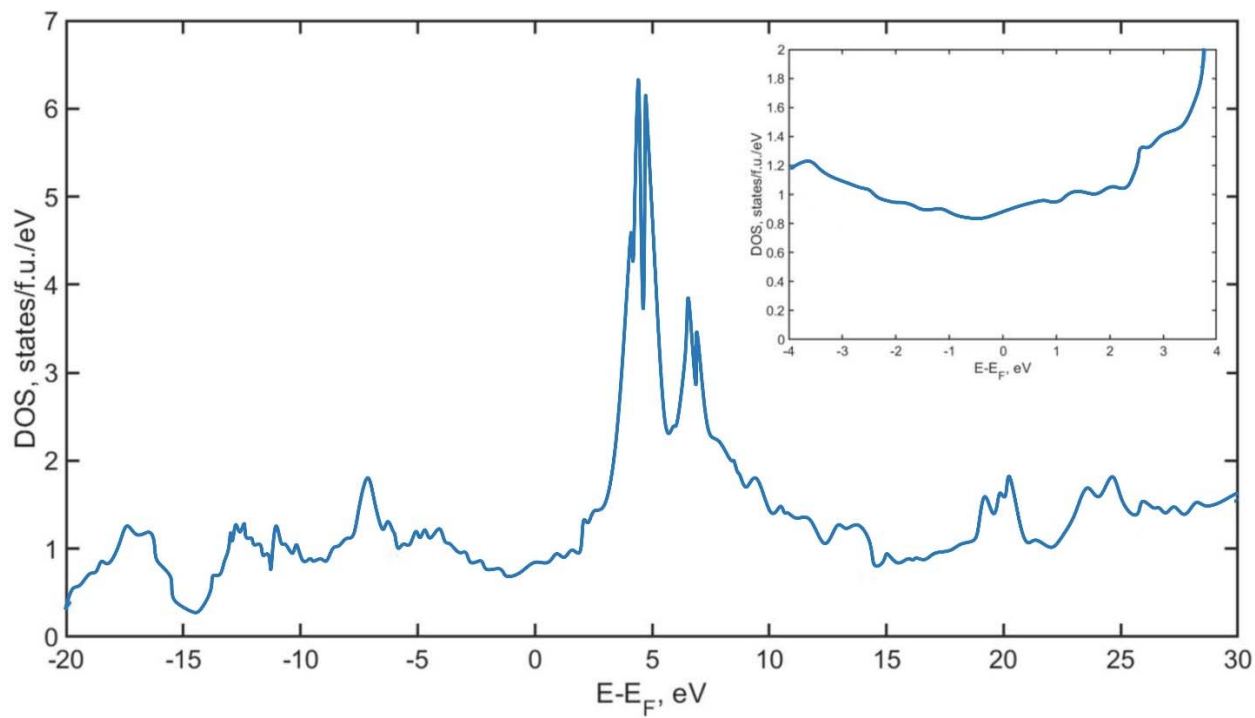

Fig. S8. Electronic density of states of $Fm\bar{3}m$-LaH$_{10}$ at 200 GPa.

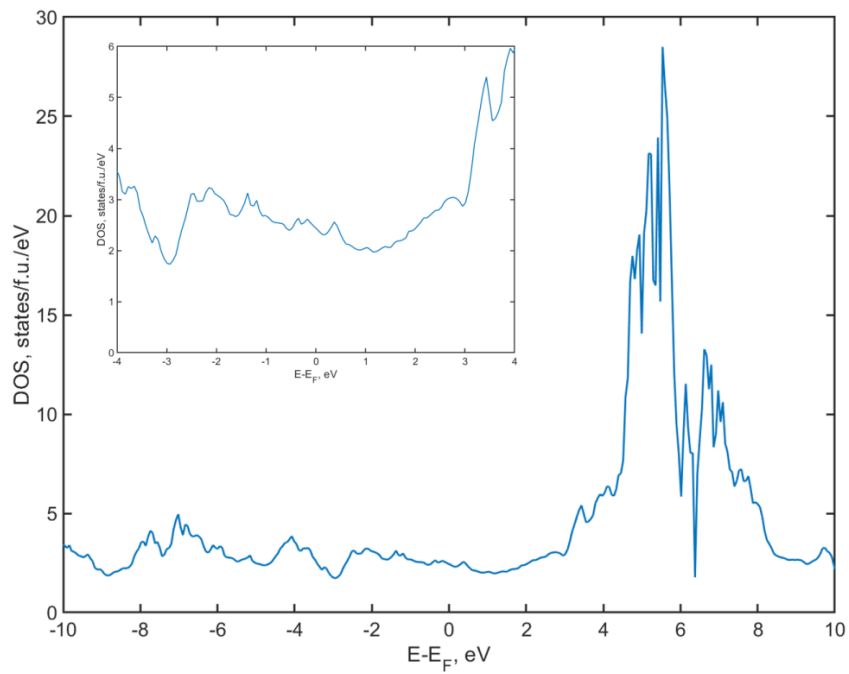

Fig. S9. Electronic density of states of $R\bar{3}m$-LaH$_{10}$ at 165 GPa.

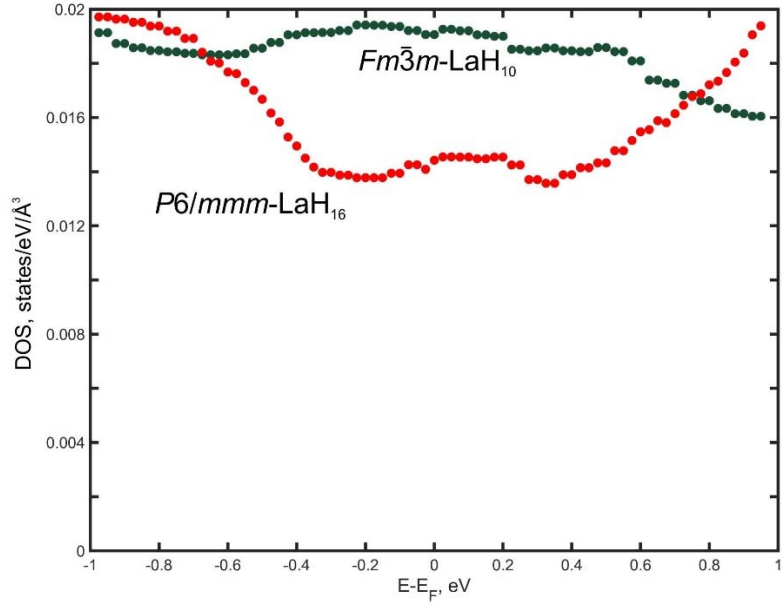

Fig. S10. Electronic density of states around the Fermi level in $Fm\bar{3}m$-LaH$_{10}$ and $P6/mmm$-LaH$_{16}$ at 200 GPa.

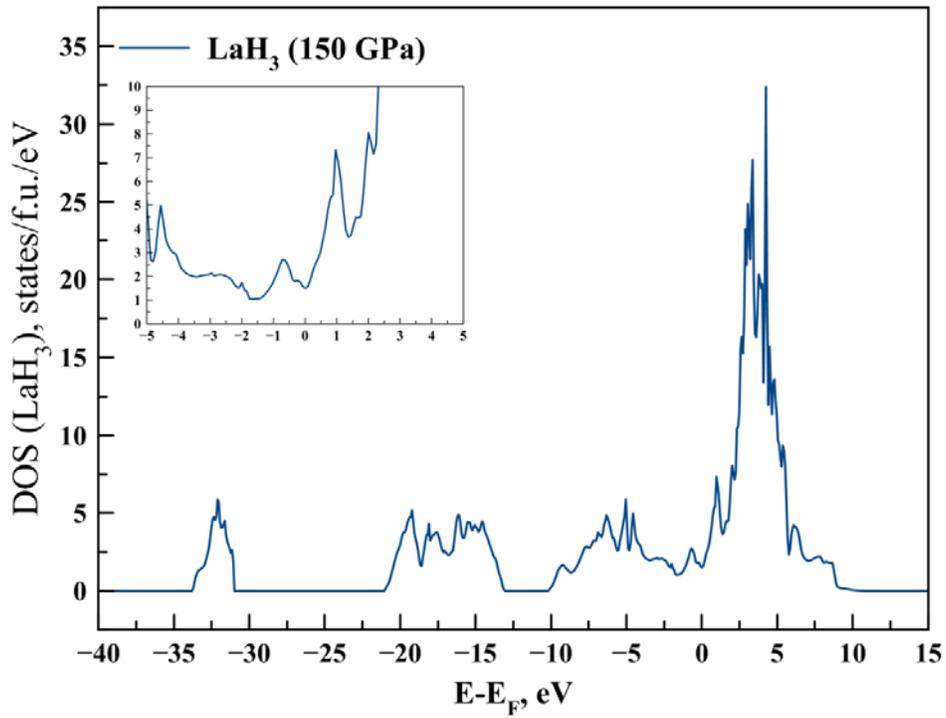

Fig. S11. Electronic density of states of $Cmcm$-LaH$_3$ at 150 GPa.

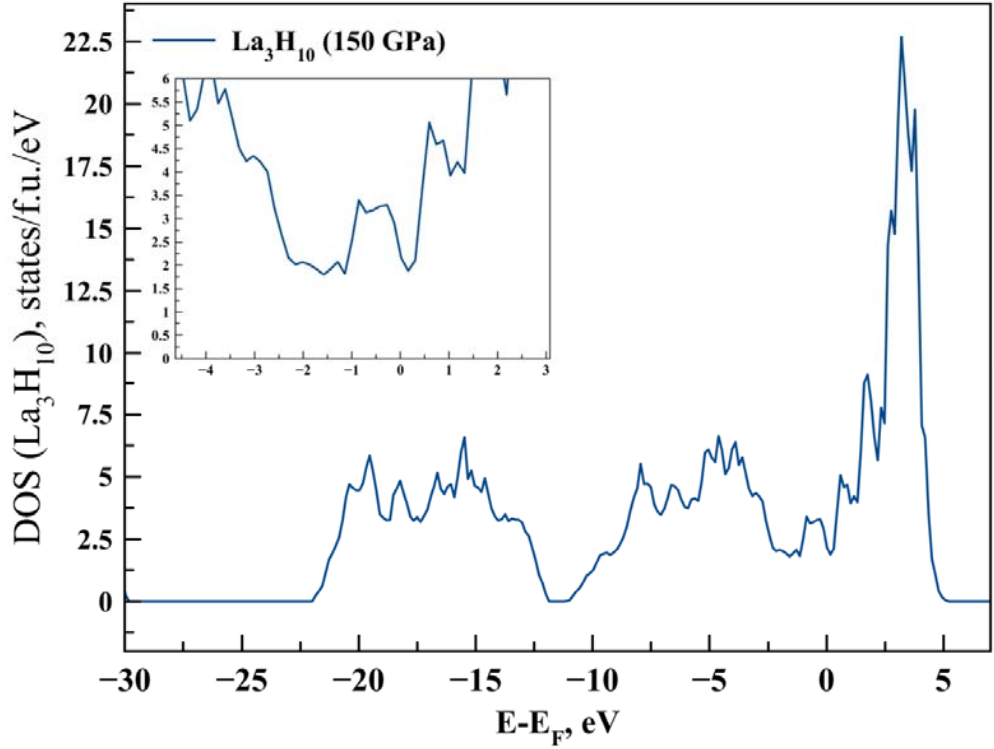

Fig. S12. Electronic density of states of *Cmmm*-La$_3$H$_{10}$ at 150 GPa.

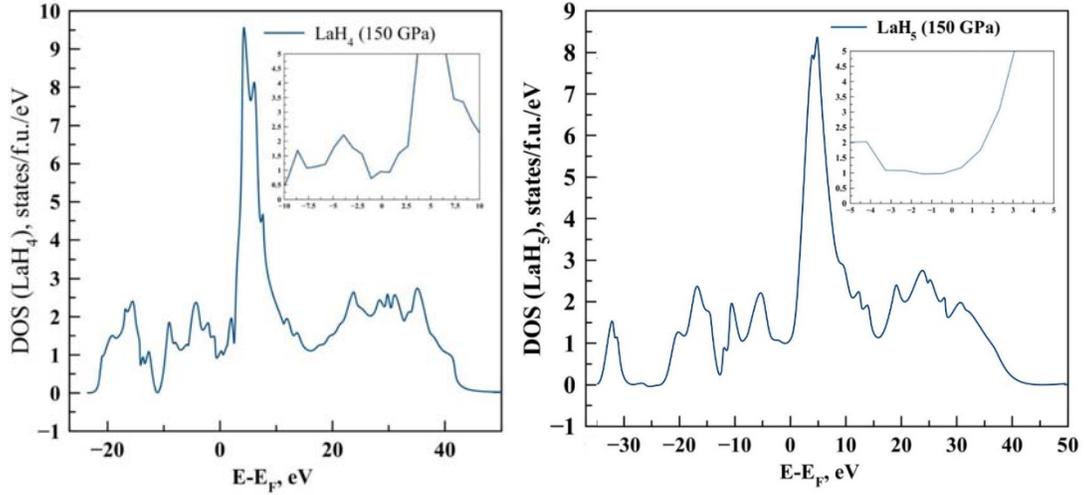

Fig. S13. Electronic density of states of $I4/mmm$-LaH$_4$ and $P\bar{1}$-LaH$_5$ at 150 GPa.

# Eliashberg spectral functions

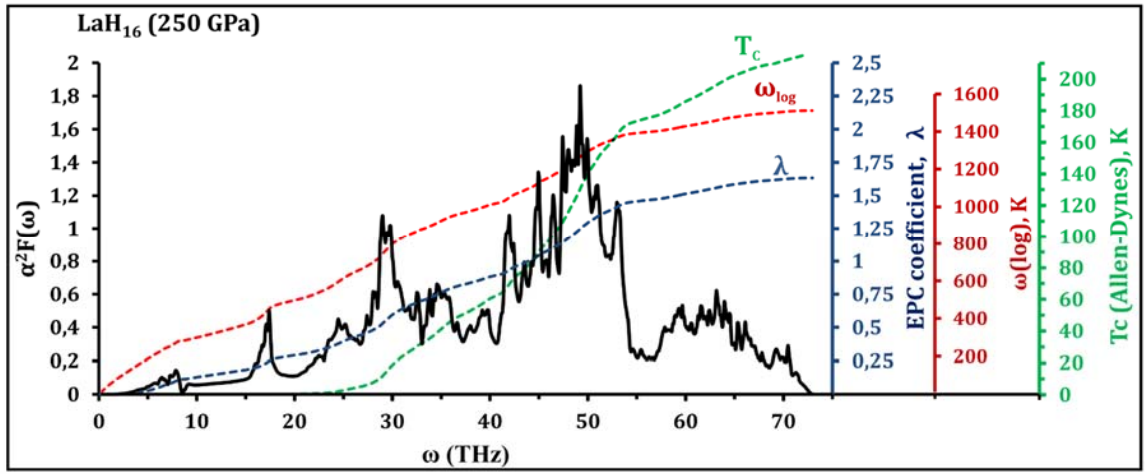

Fig. S14. Eliashberg function $\alpha^2F(\omega)$ (black), $\omega_{log}$ (red), EPC coefficient $\lambda$ (blue), critical transition temperature $T_C$ (green) of $P6/mmm$-LaH$_{16}$ at 250 GPa.

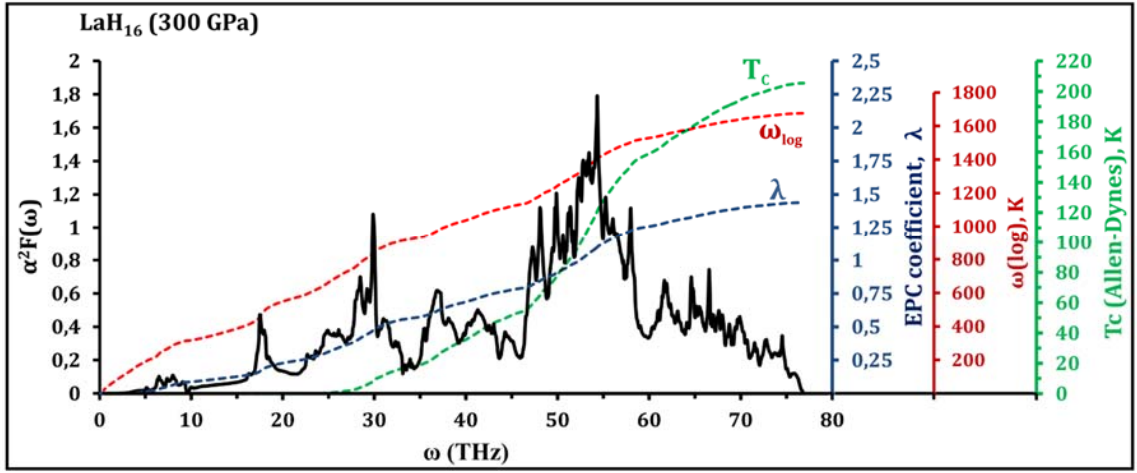

Fig. S15. Eliashberg function $\alpha^2F(\omega)$ (black), $\omega_{log}$ (red), EPC coefficient $\lambda$ (blue), critical transition temperature $T_C$ (green) of $P6/mmm$-LaH$_{16}$ at 300 GPa.

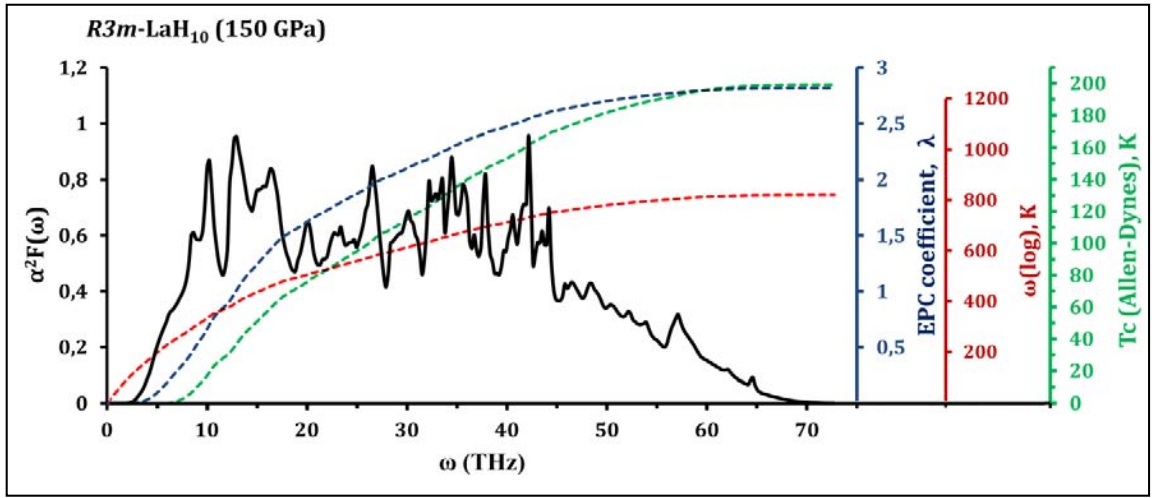

Fig. S16. Eliashberg function $\alpha^2F(\omega)$ (black), $\omega_{log}$ (red), EPC coefficient $\lambda$ (blue), critical transition temperature $T_C$ (green) of $R\bar{3}m$-LaH$_{10}$ at 150 GPa.

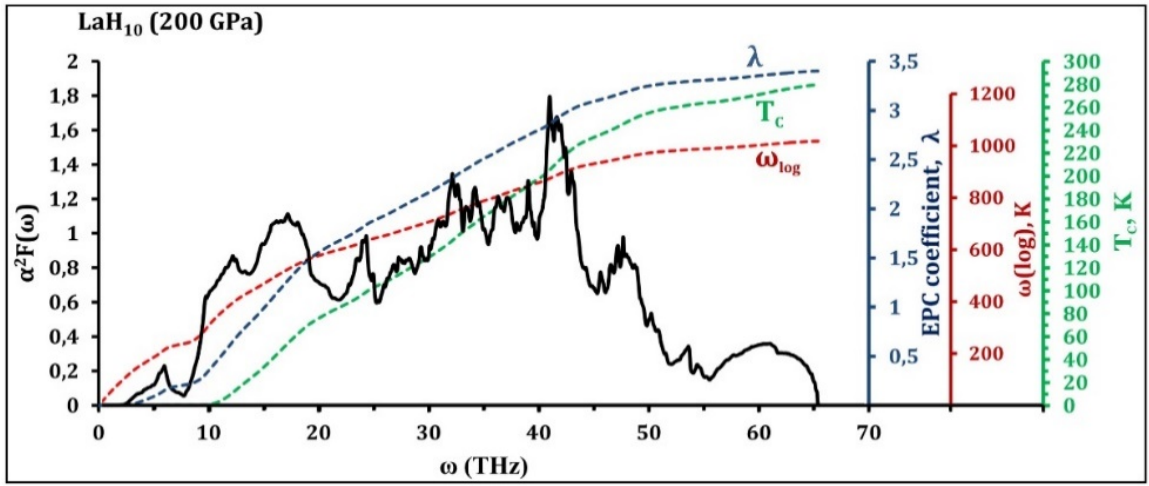

Fig. S17. Eliashberg $\alpha^2F(\omega)$ function (black curve), and cumulative $\omega_{log}$ (red), EPC coefficient $\lambda$ (blue), critical transition temperature $T_C$ (green) of $Fm\bar{3}m$-LaH$_{10}$ at 200 GPa.

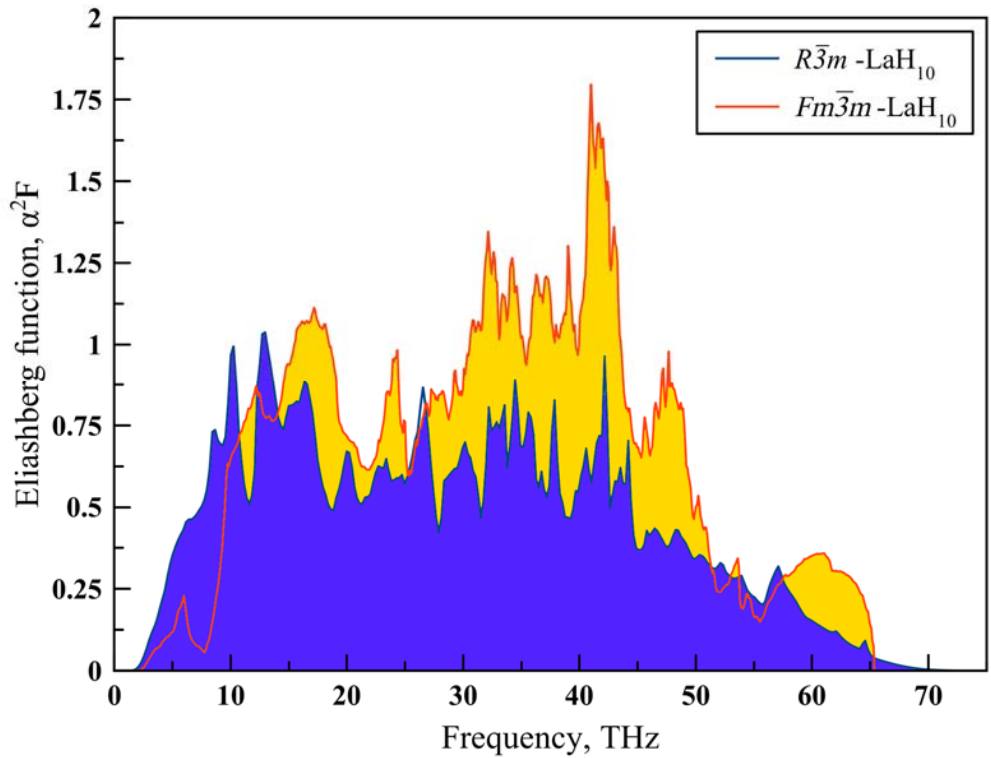

Fig. S18. Eliashberg functions $\alpha^2F(\omega)$ of $R\bar{3}m$-LaH$_{10}$ at 150 GPa (blue) and $Fm\bar{3}m$-LaH$_{10}$ at 200 GPa (yellow).

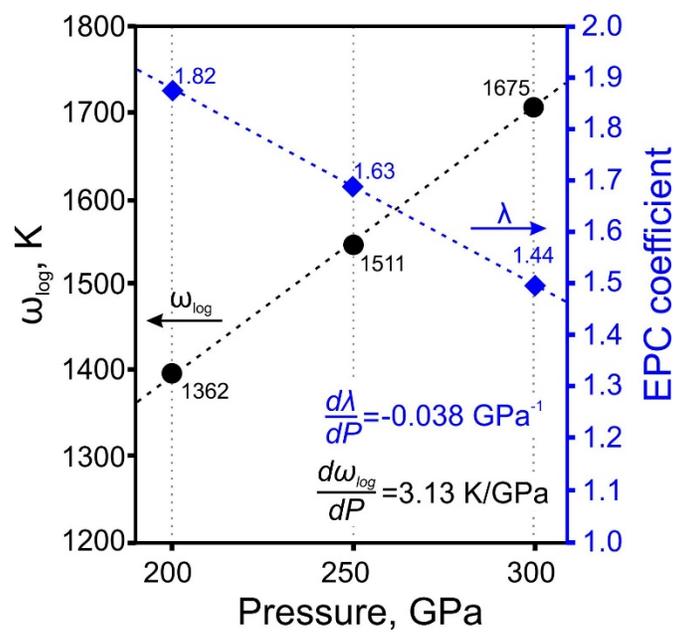

Fig. S19. Dependence of EPC coefficient and $\omega_{log}$ on pressure for $LaH_{16}$.

# Stability of LaH₁₀ polymorphs

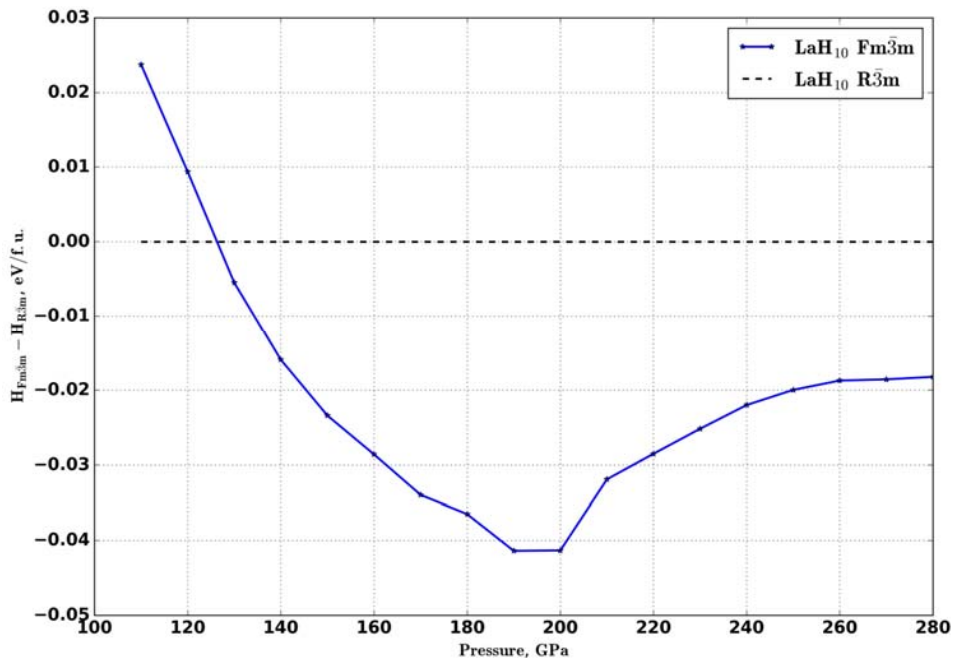

Fig. S20. Enthalpy difference between $Fm\bar{3}m$-LaH$_{10}$ and $R\bar{3}m$-LaH$_{10}$. Undoubtedly, the transition $Fm\bar{3}m \rightarrow R\bar{3}m$ plays an important role in the measured record superconductivity of LaH$_{10}$. Such structural transitions accompany superconductivity in H$_3$S ($C2/c \rightarrow R\bar{3}m \rightarrow Im\bar{3}m$) [24], in Nb$_3$Al, V$_3$Ga [25] and in other well-known superconductors.